\documentclass[prb,twocolumn]{revtex4}
\usepackage{amssymb}
\usepackage{amsmath}
\usepackage{graphicx}
\usepackage{dcolumn}
\usepackage{bm}
\usepackage[bookmarks = true, pdfpagemode = None, pdfstartview = FitH,colorlinks = true,citecolor = black, linkcolor = black, urlcolor = black]{hyperref}
\usepackage{url}

\newcommand{\bra}[1]{\left\langle{#1}\right\vert}
\newcommand{\ket}[1]{\left\vert{#1}\right\rangle}
\newcommand{\iqp}{\left|I_q^p\right|}
\newcommand{\Phiqx}{\Phi^x_q}

\newcommand{\Phiccjjx}{\Phi^x_{\text{ccjj}}}

\newcommand{\Philatchx}{\Phi^x_{\text{latch}}}
\newcommand{\Phirox}{\Phi^x_{\text{ro}}}

\newcommand{\hamisg}{{\cal H}_{\text{ISG}}}
\newcommand{\hamqsg}{{\cal H}_{\text{QSG}}}
\newcommand{\ham}{{\cal H}_0}

\begin{document}

\title{Experimental Investigation of an Eight-Qubit Unit Cell\\ in a Superconducting Optimization Processor}
\author{R. Harris}
\author{M.W. Johnson}
\author{T. Lanting}
\author{A.J. Berkley}
\author{J. Johansson}
\author{P. Bunyk}
\author{E. Tolkacheva}
\author{E. Ladizinsky}
\author{N. Ladizinsky}
\author{T. Oh}
\author{F. Cioata}
\author{I. Perminov}
\author{P. Spear}
\author{C. Enderud}
\author{C. Rich}
\author{S. Uchaikin}
\author{M.C. Thom}
\author{E.M. Chapple}
\author{J. Wang}
\author{B. Wilson}
\author{M.H.S. Amin}
\author{N. Dickson}
\author{K. Karimi}
\author{B. Macready}
\author{C.J.S. Truncik}
\author{G. Rose}
\affiliation{D-Wave Systems Inc., 4401 Still Creek Drive, Burnaby BC, Canada, V5C 6G9}
\date{\today}

\begin{abstract}
A superconducting chip containing a regular array of flux qubits, tunable interqubit inductive couplers, an XY-addressable readout system, on-chip programmable magnetic memory, and a sparse network of analog control lines has been studied.  The architecture of the chip and the infrastructure used to control it were designed to facilitate the implementation of an adiabatic quantum optimization algorithm.  The performance of an eight-qubit unit cell on this chip has been characterized by measuring its success in solving a large set of random Ising spin glass problem instances as a function of temperature.  The experimental data are consistent with the predictions of a quantum mechanical model of an eight-qubit system coupled to a thermal environment.  These results highlight many of the key practical challenges that we have overcome and those that lie ahead in the quest to realize a functional large scale adiabatic quantum information processor.
\end{abstract}

\maketitle

\section{Introduction}

Of the known paradigms of quantum computing, those related to quantum annealing (QA) \cite{Kadowaki,Farhi1,Santoro,SantoroReview,Lidar1} are unique in that they leverage what ought to be a robust natural phenomenon - the tendency for physical systems to seek and remain in a low energy configuration.  These methods have received considerable theoretical attention of late with claims and counter-claims concerning their ultimate utility \cite{Altshuler1,Farhi2,Altshuler2,AminAndChoi,Young1}.  In contrast, they have received relatively little experimental attention.  This is a significant concern as data are needed to help clarify the impact of environmental noise upon such approaches to quantum computing \cite{AoRammer,Childs,Sarandy,Roland,Ashab,Tiersch,Fubini,OHara,Crazy}.  Experiments have been performed using nuclear magnetic resonance in molecules \cite{Chuang},  but the prospects for scaling this approach to larger systems are limited.  At least one proposal for an adiabatic quantum optimization (AQO) processor based upon superconducting flux qubits exists \cite{Kaminsky1,Kaminsky2} and rudimentary experiments have been performed upon related small-scale devices with fixed qubit and coupler parameters \cite{IPHT3QProposal,IPHT4Q,IPHTClassicalCircuit}.  Experimental investigations of QA in large-scale systems have only ever been performed upon solid state samples in which the experimenter has no control over the individual couplings between Ising spins \cite{Brooke1,Brooke2}.  Consequently, there are many open questions regarding how to implement a large-scale programmable QA information processor in practice.  

The purpose of this article is threefold:  First, we address some of the practical questions regarding how to design a scalable superconducting AQO information processor.  The answers to these questions will then serve as a motivation for the architecture of the device that we have fabricated.  Second, we present experimental results from a unit cell on one such chip consisting of eight flux qubits \cite{CCJJ}, sixteen in situ tunable inductive interqubit couplers \cite{CJJCoupler}, an XY-addressable high fidelity readout architecture \cite{QFP}, an array of in situ programmable flux storage devices addressed by a single flux quantum demultiplexing circuit \cite{PMM}, and a sparse network of analog control lines.  The data demonstrate that the unit cell can be used as a computer for solving Ising spin glass problems.  Third, we compare experimental data to the results of numerical simulations in order to highlight the fact that, when run very slowly with respect to the adiabatic limit, the performance of the unit cell is influenced by its tendency to thermalize to an environment.

This article is organized as follows: Section \ref{sec:theory} contains a mathematical description of how the physics of an Ising spin glass can be mapped onto superconducting hardware composed of rf-SQUID flux qubits and couplers.  Section \ref{sec:device} contains a brief review of the hardware that has been been fabricated and key calibration data from one of the eight-qubit unit cells.  This particular unit cell was used to solve a large number of Ising spin glass problem instances and the results have been summarized in Section \ref{sec:experiment}.  A dynamical model of the unit cell coupled to an environment in thermal equilibrium is introduced in Section \ref{sec:simulation}. 
Section \ref{sec:discussion} contains a brief discussion of the key results of this article and presents a series of important open questions that have been motivated by this work.  The conclusions are summarized in Section \ref{sec:conclusions}.  For convenience, we have provided a glossary of abbreviations used in this article in Appendix \ref{Glossary}.  A table of specific example problem instance settings referred to in Sections \ref{sec:experiment} and \ref{sec:simulation} has been placed in Appendix \ref{InstanceSettings}.

\section{Mapping Adiabatic Quantum Optimization onto Hardware}
\label{sec:theory}

We begin by mapping a particular class of optimization problems onto a scalable processor architecture that uses superconducting flux qubits.  Significant emphasis will be placed upon minimizing the number of unique {\it time-dependent} control signals, thus making efficient use of the limited number of externally supplied biases that can possibly be routed to such a processor.  On the other hand, it will be assumed that unique {\it time-independent} control signals can be generated locally on chip using a scalable form of programmable magnetic memory (PMM) that has been described in detail elsewhere \cite{PMM}.

Let there be an optimization problem of the form
\begin{equation}
\label{eqn:QUBO}
E(\vec{s})=-\sum_{i}h_is_i+\sum_{i,j>i}K_{ij}s_is_j \; ,
\end{equation}

\noindent where $s_i=\pm1$, $-1\leq h_i\leq +1$, and $-1\leq K_{ij} \leq +1$.  For any given set of $h_i$ and $K_{ij}$ there exists at least one optimal solution $\vec{s}_{\text{gs}}$ that minimizes the objective $E$.  Finding $\vec{s}_{\text{gs}}$ for such a system of coupled variables can be NP-hard \cite{BarahonaNP}.  Equation (\ref{eqn:QUBO}) can be mapped onto an Ising spin glass (ISG) Hamiltonian \cite{Barahona}
\begin{equation}\label{eqn:isg}
\frac{\hamisg}{E_0} = -\sum_i h_i \sigma_z^{(i)}+\sum_{i,j>i} K_{ij} \sigma_z^{(i)} \sigma_z^{(j)} \; ,
\end{equation}

\noindent where $\sigma_{x(z)}^{(i)}$ are Pauli matrices acting upon spin $i$ and $E_0$ is a convenient energy.  When cast in this form, $\ket{\vec{s}_{\text{gs}}}$ represents the ground state of the ISG.  The objective is to design a physical system that reliably finds $\ket{\vec{s}_{\text{gs}}}$.

One possible means of finding $\ket{\vec{s}_{\text{gs}}}$ is via the adiabatic method described in Ref.~\onlinecite{Farhi1}.  In an idealized picture of AQO, the Hamiltonian of the system at arbitrary time $t$ can be expressed as that of a quantum Ising spin glass (QSG).  In such a system, there are pairwise magnetic interactions between spins $i$ and $j$ $\propto K_{ij}$ and each spin $i$ is subjected to a local longitudinal magnetic field $\propto h_i$ and a global transverse magnetic field $\propto\Gamma(t)$:
\begin{equation}
\label{eqn:Hqisg}
\frac{\hamqsg (t)}{E_0(t)} =-\sum_i h_i \sigma_z^{(i)}+\sum_{i,j>i} K_{ij} \sigma_z^{(i)} \sigma_z^{(j)}-\Gamma(t)\sum_i\sigma_x^{(i)} \; .
\end{equation}

\noindent  One finds the lowest energy solution to an optimization problem encoded in the quantities $h_i$ and $K_{ij}$ by guiding the physical system through an evolution path described by $\Gamma(t)$ subject to the following constraints:
\begin{eqnarray*}
\Gamma(0) & \gg & h_i,K_{ij}\; , \\
\Gamma(t_f) & \ll & h_i,K_{ij} \; ,
\end{eqnarray*}

\noindent where $t_f>0$ is the run time of the algorithm.  Here, it is assumed that the physical system readily reaches its ground state $\ket{\Pi_0}$ at $t=0$ and, in the best of circumstances, will remain in the ground state during the course of the evolution, thus yielding $\ket{\vec{s}_{\text{gs}}}$ at $t=t_f$.  Understanding the conditions under which adiabaticity is violated, thus yielding final states other than $\ket{\vec{s}_{\text{gs}}}$, is a matter of considerable debate at the moment \cite{Lidar1,Altshuler1,Farhi2,Altshuler2,Young1,AminAndChoi}.

It was recognized in Ref.~\onlinecite{Kaminsky1} that one could implement a QSG using a network of rf SQUIDs in which some are designed to fulfill the role of qubits and others to fulfill the role of tunable inductive interqubit couplers.  The physics of the two lowest lying states of an isolated flux qubit $i$ can be captured by an effective Hamiltonian
\begin{equation}
\label{eqn:Hq}
{\cal H}_i=-\frac{1}{2}\left[\epsilon_i\sigma^{(i)}_z+\Delta_q\sigma^{(i)}_x\right]\; ,
\end{equation}

\noindent where $\epsilon_i=2\iqp\left(\Phi_i^x-\Phi_i^0\right)$ and $\Delta_q$ are the bias and tunneling energy, respectively.  Here, $\iqp$ represents the magnitude of the supercurrent flowing about the rf-SQUID loop, $\Phi_i^x$ is the external flux bias applied to the qubit loop about which the supercurrent flows and $\Phi_i^0$ represents the qubit degeneracy point.  Upon comparing terms with like symmetry in Hamiltonians (\ref{eqn:Hqisg}) and (\ref{eqn:Hq}), one can see that $\Delta_q/2=E_0(t)\Gamma(t)$.  In order to implement an AQO algorithm in which $\Gamma(t)$ is altered, one must provide a means for tuning $\Delta_q$ in situ.  This can be accomplished by incorporating at least one dc-SQUID loop into an rf-SQUID body \cite{CJJ,Orlando,IBM,Delft,CCJJ}.  Since we have used the compound-compound Josephson junction (CCJJ) rf SQUID in our experiments \cite{CCJJ}, let the external flux bias applied to this device be represented by $\Phiccjjx$.  This leads to the substitution $\Delta_q\rightarrow\Delta_q(\Phiccjjx)$ in Hamiltonian (\ref{eqn:Hq}).  Rather unfortunately, $\iqp$ is also a function of $\Phiccjjx$ in any compound junction rf-SQUID flux qubit: $\iqp\rightarrow\left|I_q^p(\Phiccjjx)\right|$.  Therefore, one does not naturally obtain orthogonal control of $\epsilon_i$ and $\Delta_q$ via $\Phi_i^x$ and $\Phiccjjx$, respectively.  This drawback is by no means unique to the CCJJ design, as it occurs in all flux qubits with in situ tunable $\Delta_q$ that have been reported on in the literature to date.  This intrinsic lack of orthogonal control of the two defining parameters of the flux qubit, $\iqp$ and $\Delta_q$, has driven some of the design choices made in implementing the QSG discussed herein.

Assuming that one has a set of flux qubits for which $\left|I_q^p(\Phiccjjx)\right|$ and $\Delta_q(\Phiccjjx)$ are nominally identical, a general time-dependent Hamiltonian for a system of inductively coupled flux qubits can be written as
\begin{eqnarray}
\label{eqn:Hrfs}
\ham (t) & = & -\frac{1}{2}\sum_i\Bigl[2|I^p_q\!\left(t\right)|\left(\Phi_i^x(t)-\Phi_i^0\right)\sigma_z^{(i)} 
+ \Delta_q\!\left(t\right) \sigma_x^{(i)}\Bigr] \nonumber \\
& & +\sum_{i,j>i} M_{ij}\left(t\right)| I^p_q\!\left(t\right)|^2\sigma_z^{(i)} \sigma_z^{(j)} \; ,
\end{eqnarray}

\noindent where the time dependence of the qubit parameters $\iqp$ and $\Delta_q$ is implicitly driven by $\Phiccjjx(t)$.  In addition, there could be time-dependence in the individual qubit flux biases $\Phi^x_i(t)$ and in the interqubit effective mutual inductances $M_{ij}$ via their control biases $\Phi^x_{\text{co},ij}(t)$ \cite{CJJCoupler}.  

\begin{figure}[ht]
\includegraphics[width=3.25in]{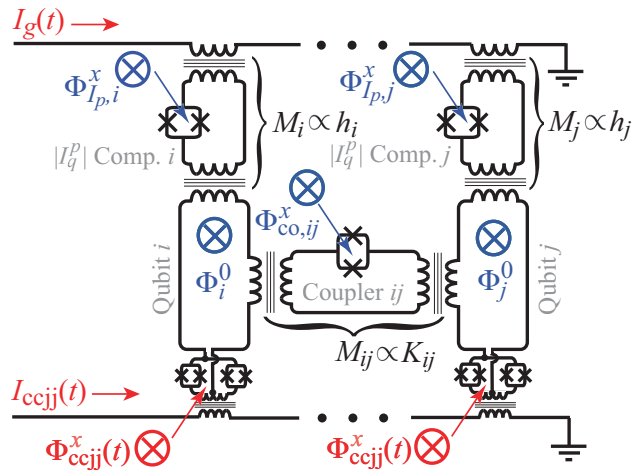}
\caption{\label{fig:IpCompensator} (Color online) Mapping of ISG problems, specified by a set of values denoted as $h_i$ and $K_{ij}$, onto superconducting hardware.  Two qubits, two $\iqp$ compensators, and one interqubit coupler are shown.  A global current bias $I_{\text{ccjj}}(t)$ provides the fluxes $\Phiccjjx(t)$ that drive the annealing process to multiple CCJJ rf SQUID flux qubits.  Interqubit coupling is mediated by tunable mutual inductances $M_{ij}\propto K_{ij}$ that are controlled by static fluxes $\Phi^x_{\text{co},ij}$.  Qubit bodies subjected to the sum of static flux biases $\Phi^0_i$ and time-dependent flux biases driven by a global current bias $I_g(t)$.  The latter signals are mediated to each qubit via tunable mutual inductances $M_i\propto h_i$ that are controlled by static fluxes $\Phi^x_{I_p,i}$.}
\end{figure}

One of the virtues of the AQO algorithm is its simplicity, which provides some significant advantages when considering how to implement a processor.  Note that Hamiltonian (\ref{eqn:Hrfs}) is a function of a {\it global} bias $\Phiccjjx(t)$ to which all qubits are uniformly subjected.  This is desirable for two reasons: First, Hamiltonian (\ref{eqn:Hqisg}) specifies that the transverse field $\Delta_q(t)/2=E_0(t)\Gamma(t)$ be uniform, which then corresponds to all $\Delta_q(\Phiccjjx)$ being identical.  As a corollary, this renders all $\left|I^p_q(\Phiccjjx)\right|$ identical.  Second, one can provide uniform $\Phiccjjx$ to multiple qubits using a single global current bias line, as opposed to one bias line per qubit.  This scenario is depicted in Fig.~\ref{fig:IpCompensator}, in which the current bias $I_{\text{ccjj}}(t)$ drives the fluxes $\Phiccjjx$ in a multitude of qubits.  In doing so, one substantially reduces the number of external biases that must be applied to the processor, thus significantly improving the scaling of this particular architecture.  Such design choices are most likely necessary in order to realize multiqubit processors that contain more than 10's of qubits. 

Given the ideal Hamiltonian (\ref{eqn:Hqisg}) and the flux qubit-based device Hamiltonian (\ref{eqn:Hrfs}), one must now map the problem specification, embodied by $h_i$ and $K_{ij}$ in Eq.~(\ref{eqn:QUBO}), onto external control parameters.  This can be accomplished by scaling Hamiltonian (\ref{eqn:Hrfs}) and then performing a term-by-term comparison to Hamiltonian (\ref{eqn:Hqisg}).  A convenient scaling factor is the interqubit coupling energy when a coupler is set to provide maximum antiferromagnetic (AFM) coupling, $M_{ij}\left(\Phi^x_{\text{co},ij}\right)=M_{\text{AFM}}$.  Define this energy scale as $J_{\text{AFM}}(t)\equiv M_{\text{AFM}}\left|I_q^p(t)\right|^2$.  Rearranging Hamiltonian (\ref{eqn:Hrfs}) yields
\begin{subequations}
\begin{equation}
\label{eqn:Hrfsrearranged}
\frac{\ham (t)}{J_{\text{AFM}}(t)} = -\sum_i h_i \sigma_z^{(i)} +\sum_{i,j>i} K_{ij} \sigma_z^{(i)} \sigma_z^{(j)}-\Gamma(t)\sum_i\sigma_x^{(i)} \; ,
\end{equation}

\begin{equation}
\label{eqn:hdefn}
h_i=\frac{|I_q^p(t)|\left(\Phi_i^x(t)-\Phi_i^0\right)}{M_{\text{AFM}}|I_q^p(t)|^2}=\frac{\Phi_i^x(t)-\Phi_i^0}{M_{\text{AFM}}|I_q^p(t)|} \, ,
\end{equation}

\begin{equation}
\label{eqn:Kdefn}
K_{ij}=\frac{M_{ij}(t)|I_q^p(t)|^2}{M_{\text{AFM}}|I_q^p(t)|^2}=\frac{M_{ij}(t)}{M_{\text{AFM}}} \; ,
\end{equation}

\begin{equation}
\label{eqn:GammaDefn}
\Gamma(t)=\frac{\Delta_q(t)}{2J_{\text{AFM}}(t)} \; .
\end{equation}
\end{subequations}

In order to solve a particular optimization problem, $h_i$ and $K_{ij}$ must be {\it time-independent}.  According to Eq.~(\ref{eqn:Kdefn}), one must hold all $M_{ij}$ constant during operation.  This is convenient as it obviates the application of individually tailored time-dependent flux bias signals to each interqubit coupler.  Rather, one need only apply a static control signal $\Phi^x_{\text{co},ij}$ to each coupler, as depicted in Fig.~\ref{fig:IpCompensator}, that can be provided by PMM.  On the other hand, according to Eq.~(\ref{eqn:hdefn}), one must apply {\it time-dependent} qubit flux biases of the form
\begin{equation}
\label{eqn:hwaveform}
\Phi_i^x(t)=\Phi_i^0+h_i \times M_{\text{AFM}}|I_q^p(t)|
\end{equation}

\noindent to render $h_i$ time-independent.  Thus, it is necessary to provide a custom-tailored time-dependent control signal plus a static offset to every qubit.  The static component $\Phi_i^0$ can be provided by PMM.  As for the time-dependent component, providing these signals with one external bias per qubit would not constitute a scalable approach for building a multiqubit processor.  Rather, one can take advantage of the fact that, according to Eq.~(\ref{eqn:hwaveform}), all qubits must receive a control signal with the {\it same} time-dependent shape but with custom-tailored time-independent scale factors $\propto h_i$.  A scalable architecture for providing these signals is depicted in Fig.~\ref{fig:IpCompensator} and has been further expounded upon in Ref.~\onlinecite{PMM}.  Here, a single global current bias $I_g(t)=\alpha \left|I_q^p(t)\right|$, where $\alpha$ is a convenient scale factor, is coupled to multiple qubits via in situ tunable mutual inductances of magnitude $M_i\equiv h_iM_{\text{AFM}}/\alpha$.  The very same type of device that is used to provide in situ tunable interqubit coupling \cite{CJJCoupler} can be retooled to provide coupling between flux qubits and a global bias line.  Moreover, each $M_i$ can be controlled with a static flux bias $\Phi^x_{I_p,i}$ provided by PMM.  We will hereafter be refer to this architecture as {\it persistent current} ($\iqp$) {\it compensation},  as it is a means of compensating for changing $\left|I_q^p(t)\right|$ such that the ISG problem specified by $h_i$ and $K_{ij}$ remains on target throughout annealing.

To summarize up to this point, a prescription for implementing AQO to solve ISG problems using a network of inductively coupled CCJJ rf-SQUID flux qubits has been presented.  A problem specified by a set of $h_i$ and $K_{ij}$ can be embedded in the hardware using time-independent interqubit couplings controlled by PMM and time-dependent qubit flux and CCJJ biases.  The qubit flux bias signals can be supplied using a combination of static flux offsets provided by PMM and a single global signal $I_g(t)$ that is applied to each qubit through in situ tunable couplers that are also controlled by PMM.  The CCJJ bias can, in principle, also be provided to all qubits simultaneously using a single global control signal.

\section{Device Architecture and Calibration}
\label{sec:device}

With the mapping of the AQO algorithm onto the scalable architecture completed, we now turn to a high level description of a superconducting chip whose architecture embodies that algorithm.  All of the principal components of the processor, namely the qubits\cite{CCJJ}, couplers\cite{CJJCoupler}, readout\cite{QFP}, and PMM\cite{PMM} have been described in detail in other publications.  As such, we will only provide brief summaries of the important points as pertaining to the functioning of the collective system herein.

\begin{figure}[ht]
\includegraphics[width=3.0in]{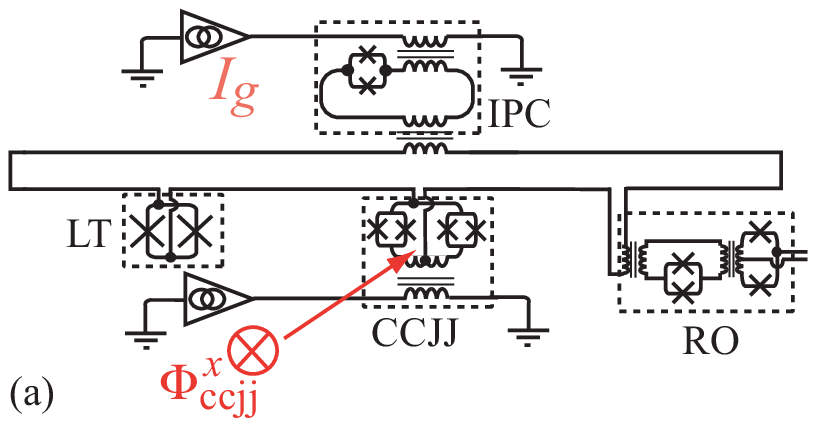}\\
\includegraphics[width=3.0in]{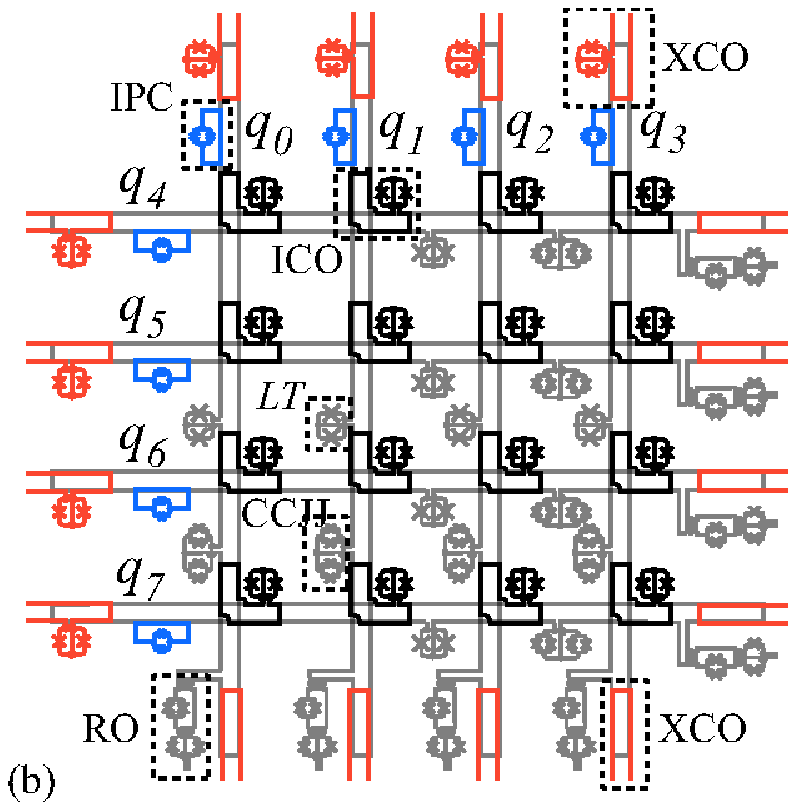}\\
\includegraphics[width=3.0in]{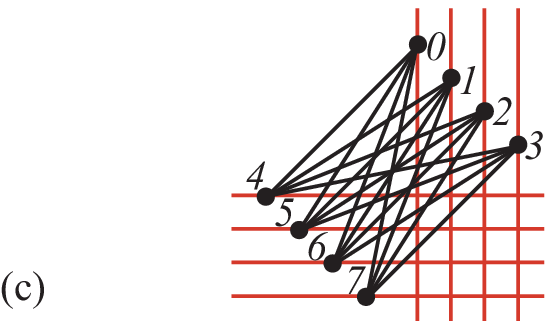}
\caption{\label{fig:DeviceSchematic} (Color online) Device schematic.  (a) A single CCJJ rf SQUID with the two time-dependent biases relevant for this study.  The flux bias $\Phiccjjx(t)$ drives the annealing process.  The global $\iqp$-compensation bias is provided by $I_g(t)$.  (b) Schematic layout of the eight-qubit unit cell.  Qubits are shown in grey.  Example readout (RO), CCJJ, $L$ tuner (LT), $\iqp$ compensator (IPC), internal coupler (ICO), and portions of external couplers (XCO) have been noted. (c) Graph representation of the hardware connectivity.  Qubits are represented by vertices (solid dots) and couplings by edges (solid lines).  Black (red) edges correspond to ICO (XCO) couplings.}
\end{figure}

As stated previously, we have incorporated CCJJ rf-SQUID flux qubits in our design \cite{CCJJ}.  A schematic of an isolated flux qubit with the two external bias controls relevant for this study is shown in Fig.~\ref{fig:DeviceSchematic}(a).  This particular qubit is robust against fabrication variations in the Josephson-junction critical currents and facilitates the homogenization of the net critical current among a population of such qubits.  This device also contains an inductance ($L$) tuner that can be used to compensate for variations of qubit inductance due to fabrication and from tuning the interqubit couplers \cite{CJJCoupler}.  To each qubit we have added an $\iqp$ compensator, as introduced in Sect.~\ref{sec:theory}.  We have provided PMM to flux bias the two minor lobes of each CCJJ \cite{CCJJ}, the $L$ tuner, the $\iqp$ compensator, and the qubit body ($\Phi_i^0$) for all qubits on the chip \cite{PMM}.

The chip that was used for this study was composed of sixteen eight-qubit unit cells that were tiled on a $4\times 4$ square grid.  To limit the scope of this article, we focus upon a single unit cell near the center of the chip.  A discussion of the complete processor, with multiple unit cells acting in concert, will be reserved for a future publication.  A schematic layout of the unit cell is shown in Fig.~\ref{fig:DeviceSchematic}(b).  The qubits are depicted as extended horizontal and vertical loops with 16 compound Josephson junction (CJJ) couplers located at the intersections of the qubits.  Additional couplers at both extrema of the qubit bodies connect them to qubits in neighboring unit cells to the left, right, top and bottom.  These latter couplers were set to zero coupling to isolate the single unit cell for this study.  Each qubit was connected to its own quantum flux parametron (QFP) enabled readout \cite{CCJJ,QFP}.  Not shown in Fig.~\ref{fig:DeviceSchematic}(b) are the PMM elements, the demultiplexing tree for addressing the PMM, or the analog bias lines used to calibrate and operate the unit cell.  

A simplified representation of the hardware connectivity is depicted in Fig.~\ref{fig:DeviceSchematic}(c).  Here, qubits and couplers correspond to the vertices and edges of a graph \cite{graph}.  As with the unit cell schematic, this eight-vertex graph can be tiled to the left, right, top, and bottom in order to generate larger graphs.  Each vertex is connected to a minimum (maximum) of 4 (6) other vertices, depending upon the number and arrangement of unit cells used to form a larger graph.  This hardware does not provide full connectivity within a population of $N$ qubits in which each physical qubit is connected to $N-1$ other physical qubits.  This limitation can potentially be overcome, at the cost of reducing the number of unique vertices $< N$, by using a ferromagnetically coupled chain of physical qubits to form a single logical qubit, as suggested in Ref.~\onlinecite{Kaminsky2}.  


The chip used in these experiments was fabricated on an oxidized Si wafer with Nb/Al/Al$_2$O$_3$/Nb trilayer junctions and four Nb wiring layers separated by planarized plasma enhanced chemical vapor deposited SiO$_{2}$.   A scanning electron micrograph of the fabrication cross-section can be found in Fig.~4(b) of Ref.~\onlinecite{CCJJ}. 
The Nb metal layers are referred to as BASE, WIRA, WIRB and WIRC, from bottom to top, respectively.  Flux qubit wiring was primarily located in WIRB and consisted of $2$-$\mu$m-wide leads arranged as an approximately $900$-$\mu$m-long differential microstrip located $200\,$nm above a ground plane in WIRA.  Coupler wiring was primarily located in WIRC, stacked on top of the qubit wiring to provide inductive coupling.  PMM flux storage loops were implemented as stacked spirals of 13-20 turns of $0.25$-$\mu$m-wide wiring with $0.25\,\mu$m separation in BASE and WIRA (WIRB).  Stored flux was picked up by one-turn washers in WIRB (WIRA) and fed into transformers for flux-biasing devices.  External control lines were mostly located in BASE and WIRA.  Resistors that were used in the PMM demultiplexing circuit were made from a TiPt layer referred to as RESI.  All of these control elements resided below a ground plane in WIRC.  The ground planes under the qubits and over the PMM/external control lines were electrically connected using extended vias in WIRB so as to form a  nearly continuous superconducting shield between the analog devices on top and the bias circuitry below.  Transformers for biasing qubits, couplers, QFPs and dc SQUIDs were enclosed in superconducting boxes with BASE and WIRC forming the top and bottom, respectively, and vertical walls formed by extended vias in WIRA and WIRB.  Minimally sized openings were placed in the vertical walls through which the bias and target device wiring passed at opposing ends of each box.  This design reduced most on-chip parasitic crosstalk to a negligible level.

\begin{figure}[ht]
\includegraphics[width=3.0in]{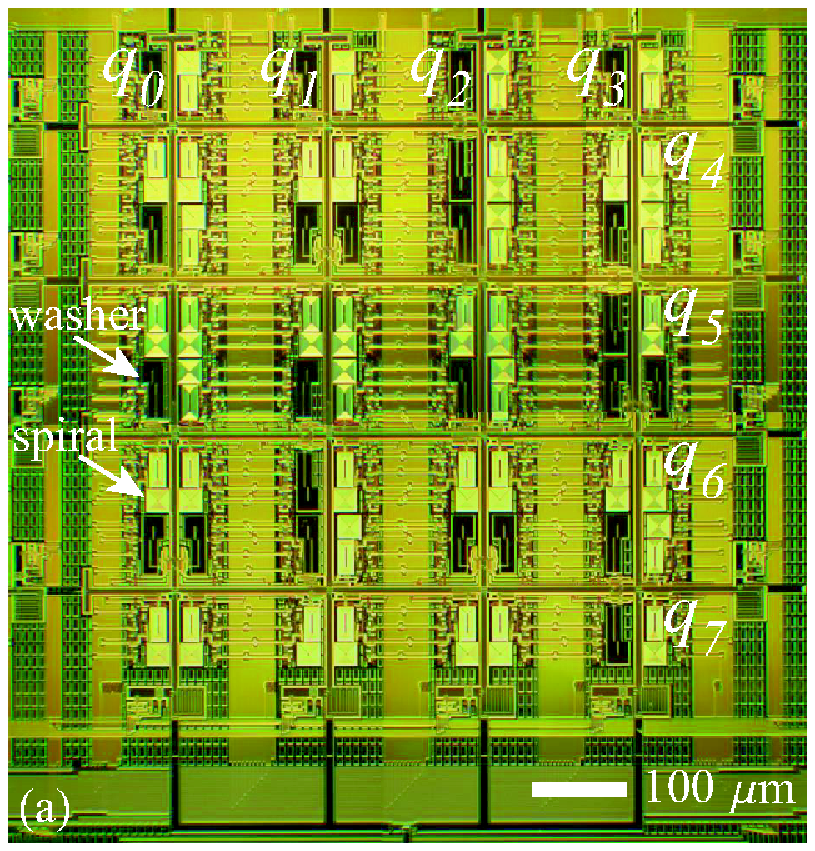}\\
\includegraphics[width=3.0in]{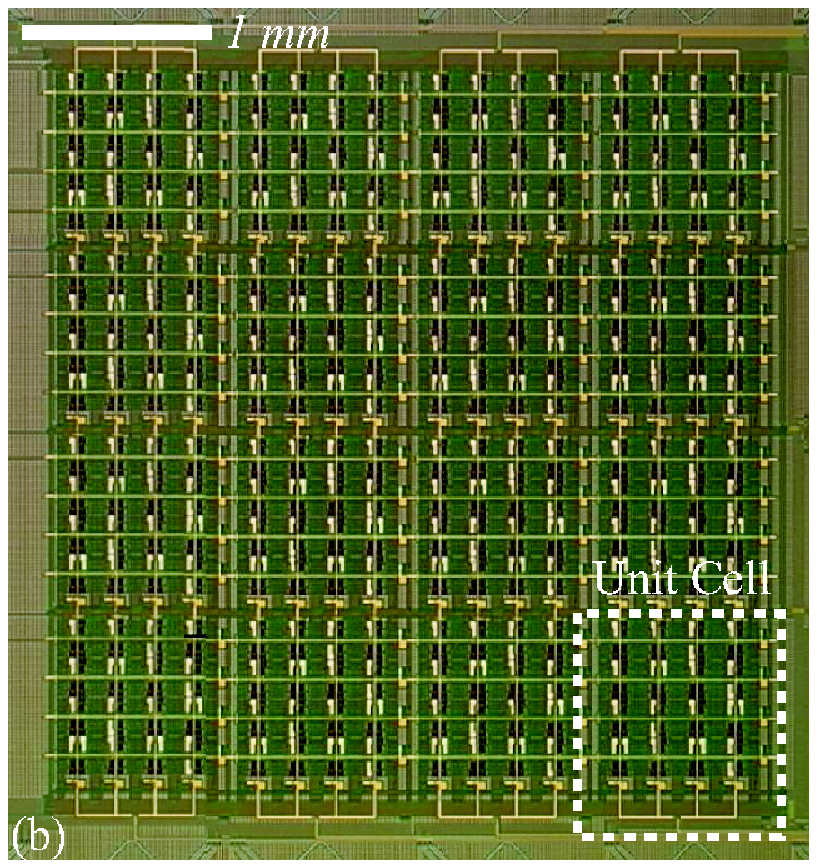}
\caption{\label{fig:processor} (Color online) 
(a) Optical image of an eight-qubit unit cell completed through the processing of WIRB.  Qubits, labeled as $q_0\rightarrow q_7$, reside within the trenches formed by ground plane in WIRA and extended vias in WIRB.  PMM elements are visible as spirals and washers between qubits.  (b) Optical image of a $4\times4$ array of eight-qubit unit cells completed through the processing of WIRB.  Unit cell shown in (a) is enclosed within the dashed box.}
\end{figure}

An optical image of a unit cell completed through the processing of WIRB is shown in Fig.~\ref{fig:processor}(a).  One can discern the trenches in which the qubit wiring resides, where the bottom is formed by ground plane in WIRA that is electrically connected to thick vertical walls that are formed by extended vias in WIRB.  A number of PMM flux storage and pickup loops are visible as spirals and washers, respectively.  The PMM circuitry is not visible in a completed chip as it resides below patches of ground plane in WIRC that are electrically connected at their perimeter to the top of the vertical walls defining the qubit trenches.  Thus the qubits (PMM) reside above (below) the nearly contiguous contiguous shielding layer in a completed chip.  Note that much of the coupler wiring is absent from this image as it resides in WIRC atop the qubit wiring in WIRB.

Tiling of the eight-qubit unit cell to make a larger processor is explicitly demonstrated in Fig.~\ref{fig:processor}(b).  One can discern a $4\times4$ array of eight-qubit unit cells.  The particular unit cell used in this study was in the third column and third row.  Outside of the field of this image are four blocks of on-chip hysteretic dc SQUIDs, with pickup loops oriented in-plane and orthogonal to the wafer, that are used as magnetometers to measure local fields during active field compensation, as well as wiring channels that run to wirebonding pads.  There are 128 qubits, 352 couplers and 128 readouts on this processor.  These analog components would require a total of 1632 unique biases to operate if no effort was made to develop a scalable control architecture.  It would be impractical to build such a processor with so many independent external biases due to circuit density and wirebonding constraints.  However, our complete processor, consisting of the analog components enumerated above and 992 PMM elements, requires only 84 pairs of differentially-driven external biases to calibrate and operate.

The chip was enclosed in a superconducting Al shield ($T_c\approx 1.2\,$K) that was inside two coaxial $\mu$-metal shields.  All three shielding layers and the sample were thermalized to the mixing chamber of a pulse tube dilution refrigerator.  The refrigerator itself was located inside two room temperature coaxial $\mu$-metal shields.  The magnetic field in the vicinity of the chip was minimized at $3.9\,$K using triple-axis Cu compensation coils located outside of the Al shield.  The array of on-chip dc-SQUID magnetometers surrounding the qubit circuitry was used to infer the magnetic field vector near the processor.  The current in the compensation coils was controlled via feedback implemented in software.  Once the magnitude of the three-dimensional magnetic field vector had been minimized, the chip was thermally cycled to $\gtrsim 9.5\,$K to expel trapped flux from the Nb chip ($T_c\approx 9.1\,$K) and then cooled to base temperature $T\approx 19\,$mK.  Thereafter, the current applied to the compensation coils was removed.  The residual magnetic field was estimated to be $\sim2\,$nT ($3\,$nT) perpendicular (parallel) to the plane of the chip.

External biases applied to the chip were provided by a custom-built 14-bit 128-channel differentially driven programmable $50\,$MHz arbitrary wave form generator.  The signals generated at room temperature were sent into the refrigerator using individually shielded twisted pair wires that were connected to a combination of lumped-element and copper powder filters at the mixing chamber.  To minimize the effect of environmental noise and since the AQO algorithm does not require resonant excitation of any element of the circuit, we chose to restrict the bandwidth of these bias lines to $5\,$MHz.  Signals were then routed onto Sn traces on a printed circuit board ($T_c\approx 3.7\,$K) and passed onto the chip via Al wirebonds.

\begin{figure}[ht]
\includegraphics[width=3.2in]{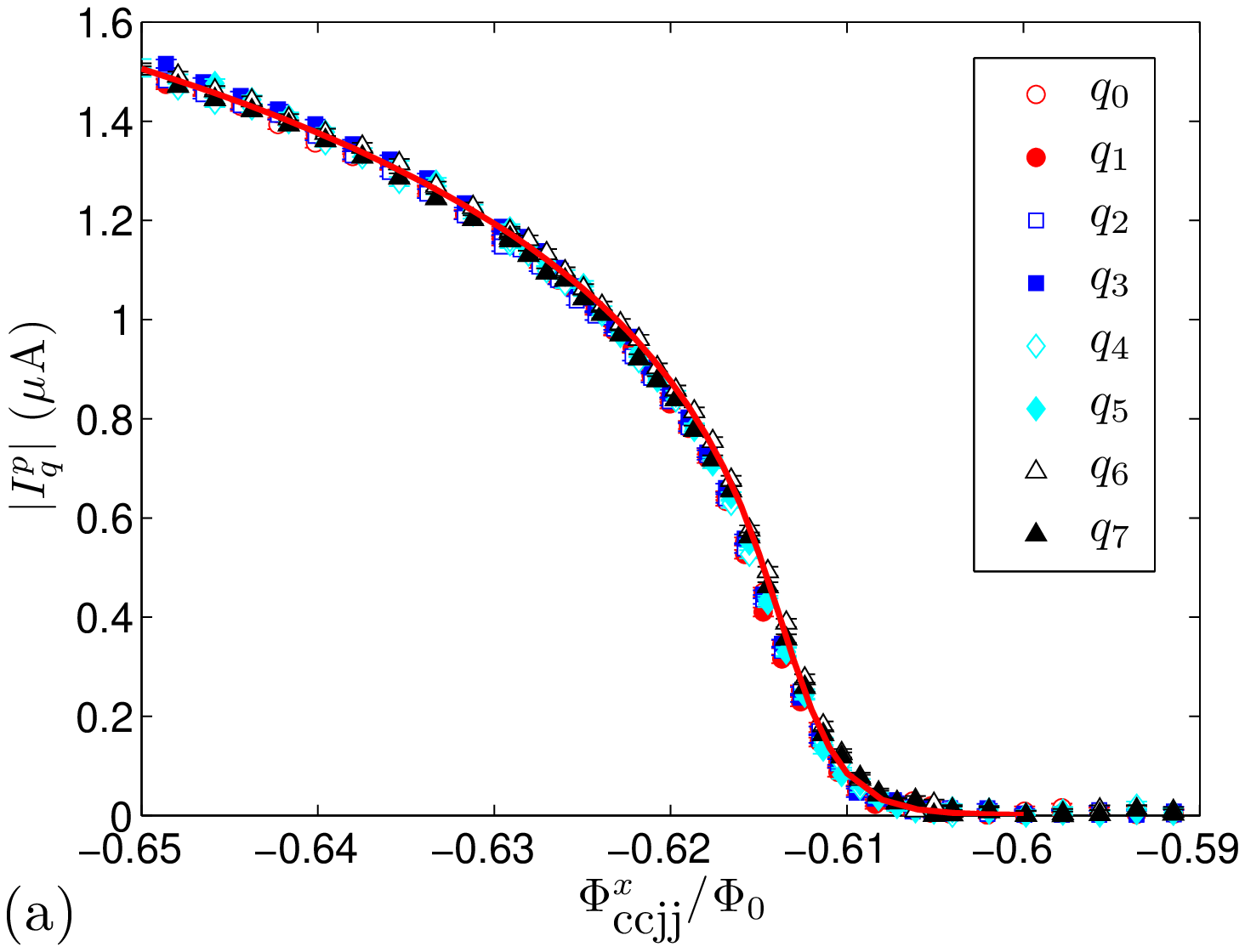} \\
\includegraphics[width=3.2in]{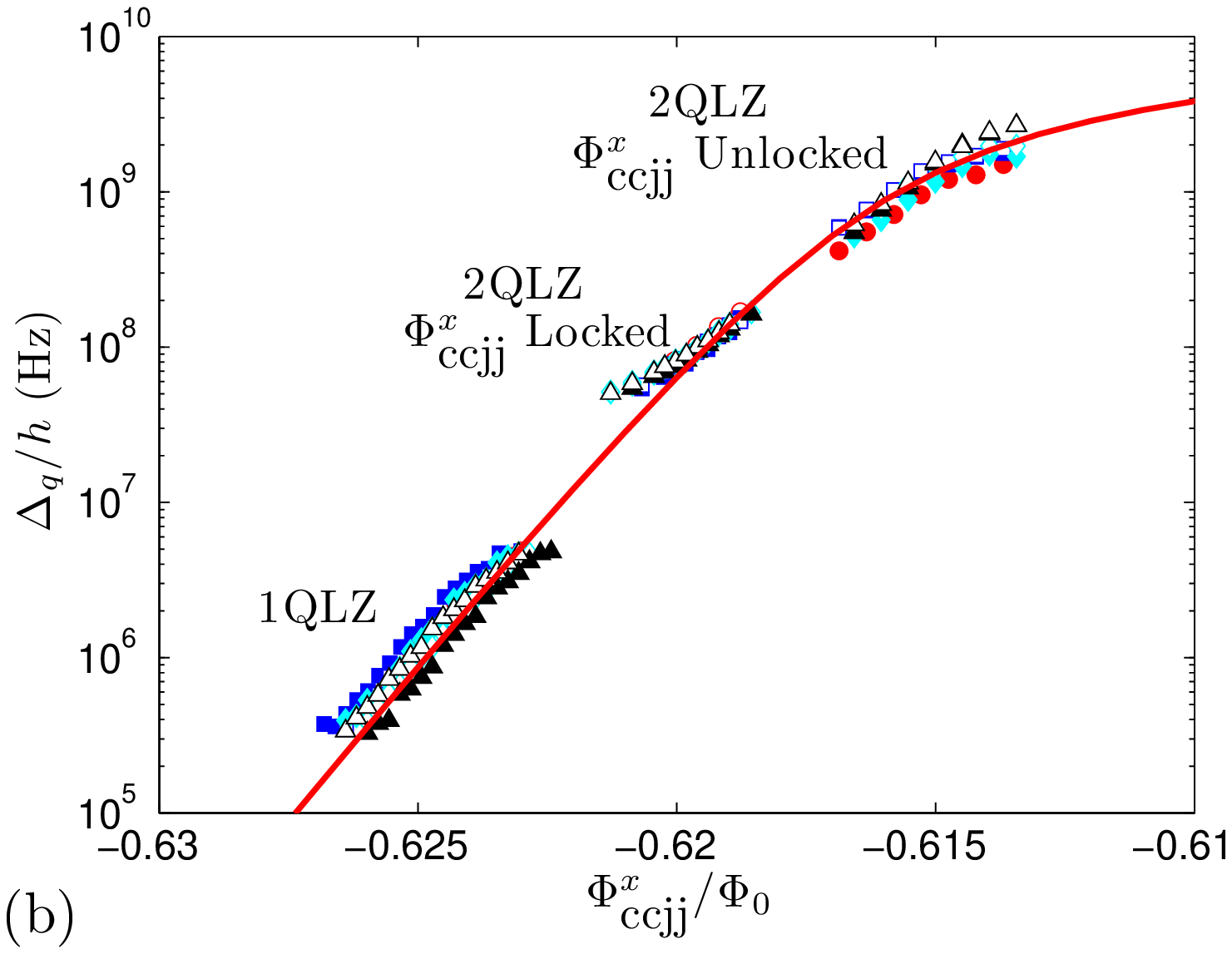}
\caption{\label{fig:QubitParameters} (Color online) Measured and predicted qubit parameters as a function of $\Phiccjjx$. (a) Magnitude of the persistent current $\iqp$. (b) Tunneling energy $\Delta_q$.  Solid curves are predictions from the CCJJ rf SQUID Hamiltonian using independently calibrated device parameters.}
\end{figure}

A summary of the flux qubit parameters that are relevant for this study is presented in Fig.~\ref{fig:QubitParameters}.  Data have been plotted versus normalized flux, where $\Phi_0\equiv h/2e$ is the flux quantum.  The reader is referred to Ref.~\onlinecite{CCJJ} for a detailed description of the CCJJ rf-SQUID flux qubit and the experimental methods used to obtain $\left|I_q^p(\Phiccjjx)\right|$.  We have obtained $\Delta_q(\Phiccjjx)$ using two methods.  First, we used the single qubit Landau-Zener (1QLZ) method \cite{LZ} to yield $2\times10^5\,\text{Hz}\lesssim\Delta_q/h\lesssim 10^7\,\text{Hz}$.  The lower and upper bound were set by experimental run time constraints and the bandwidth of our external control lines, respectively.  Second, we ran two-qubit Landau-Zener (2QLZ) experiments on pairs of qubits that were strongly coupled with $M_{ij}=-2.82\,$pH.  The Hamiltonian for a pair of qubits, subject to zero net flux bias, can be expressed as
\begin{equation}
\label{eqn:2QLZ}
{\cal H}_{ij}=-\frac{1}{2}\left[\Delta_i\sigma^{(i)}_x+\Delta_j\sigma^{(j)}_x\right]+M_{ij}\left|I_i^p\right|\left|I_j^p\right|\sigma^{(i)}_z\sigma^{(j)}_z \; .
\end{equation}

\noindent Data from a 2QLZ experiment yields a spectral gap $g$, which can be compared to that obtained by solving for the eigenenergies of Hamiltonian (\ref{eqn:2QLZ}).  The range of $g$ that could be inferred from measurements was also limited by the run time and bandwidth constraints cited above.  In the first variant of 2QLZ, we locked the target $\Phiccjjx$-biases of the qubits together.  Assuming that $\Delta_i=\Delta_j\equiv\Delta_q$, one can use Hamiltonian (\ref{eqn:Hrfsrearranged}) to solve for $\Delta_q(\Phiccjjx)$ given the independently calibrated $g(\Phiccjjx)$, $\left|I^p_i\left(\Phiccjjx\right)\right|$, and  $\left|I^p_j\left(\Phiccjjx\right)\right|$.  This procedure proved effective for extracting $5\times10^7\,\text{Hz}\lesssim\Delta_q/h\lesssim 2\times10^8\,\text{Hz}$.  In the second variant of 2QLZ, we targeted $\Phiccjjx=-0.6235\,\Phi_0$ for qubit $i$ (a bias for which we have calibrated $\Delta_i$ using the 1QLZ method)  and scanned the target $\Phiccjjx$ of qubit $j$.  We then used Hamiltonian (\ref{eqn:Hrfsrearranged}) to solve for $\Delta_j=\Delta_q(\Phiccjjx)$ given the independently calibrated $g(\Phiccjjx)$, $\Delta_i(-0.6235\,\Phi_0)$, $\left|I^p_i\left(-0.6235\,\Phi_0\right)\right|$, and  $\left|I^p_j\left(\Phiccjjx\right)\right|$.  This procedure proved effective for extracting $4\times10^8\,\text{Hz}\lesssim\Delta_q/h\lesssim 3\times10^9\,\text{Hz}$.

Prior to collecting the data shown in Fig.~\ref{fig:QubitParameters}, we had fully exercised the CCJJ minor lobe and $L$-tuner biases on each of these qubits in order to homogenize $\left|I_q^p(\Phiccjjx)\right|$ at $\Phiccjjx/\Phi_0=-0.6146$ and $-0.6325$ among the eight qubits.  Residual discrepancies between $\iqp$ and $\Delta_q$ of different qubits at arbitrary $\Phiccjjx$ were attributed to small variations in the mutual inductance between the global CCJJ current bias line and the individual qubits. 

\begin{figure}[ht]
\includegraphics[width=3.25in]{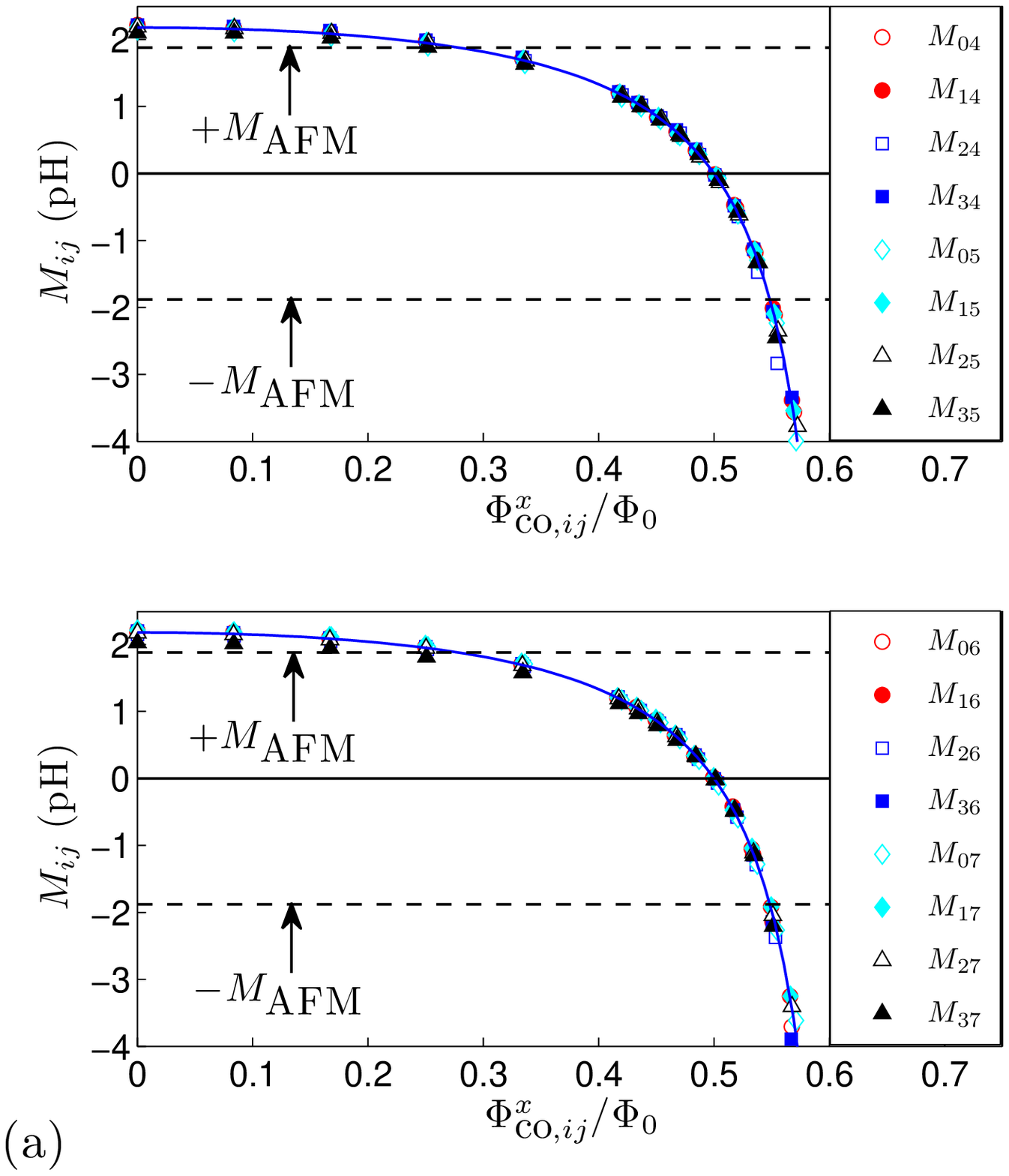}\\
\includegraphics[width=3.25in]{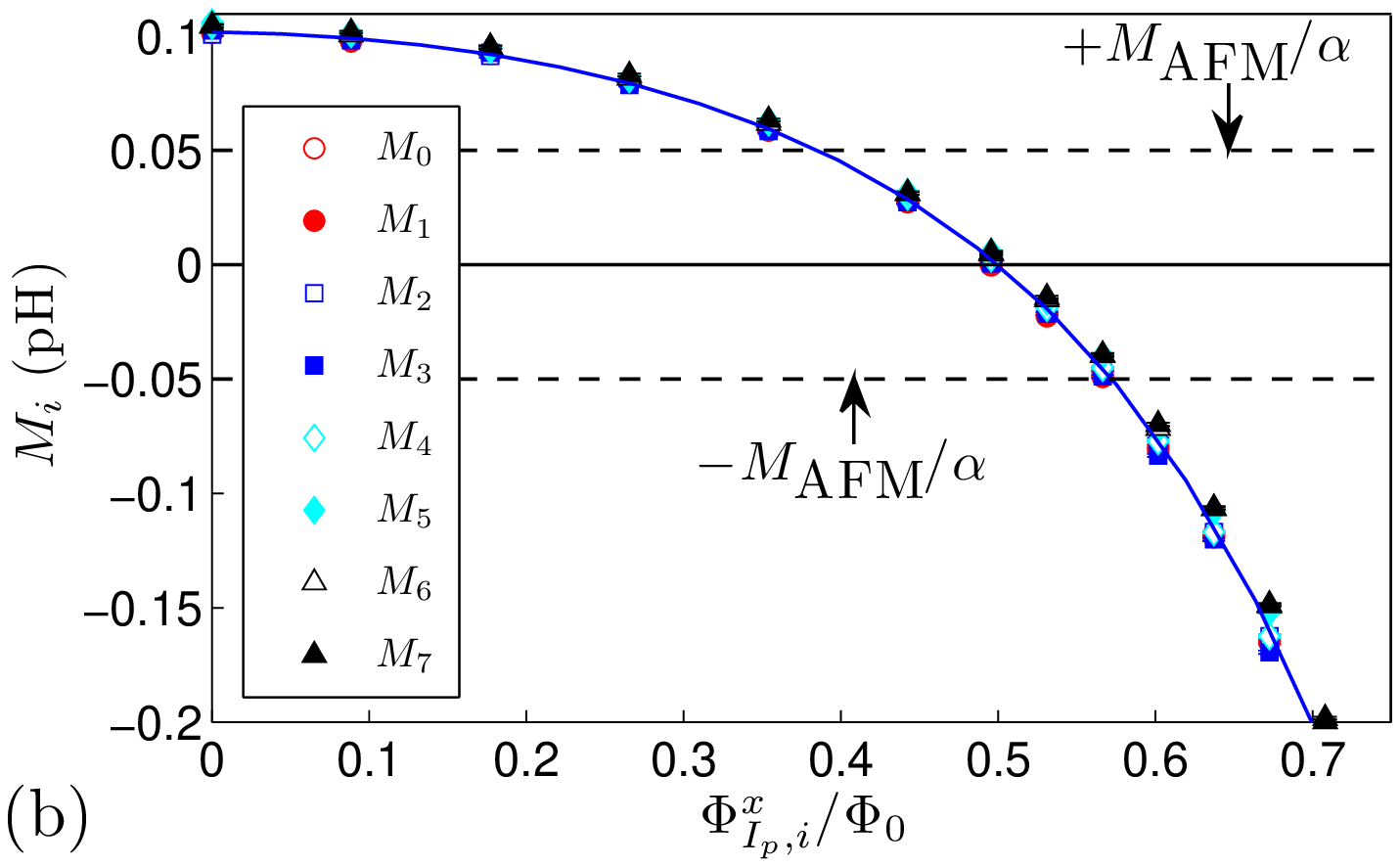}
\caption{\label{fig:CouplerAndIPC} (Color online) (a) Mutual inductance $M_{ij}$ versus coupler flux bias $\Phi^x_{\text{co},ij}$ for all sixteen interqubit couplers in the unit cell.  Solid curves are the theoretical $M_{ij}$ obtained using the mean of the individual coupler fit parameters. (b) Mutual inductance $M_{i}$ versus $\iqp$-compensator flux bias $\Phi^x_{I_p,i}$.  Solid curve is the theoretical $M_i$ obtained using the mean of the individual  $\iqp$-compensator fit parameters.}
\end{figure}

The theoretical predictions from a quantum mechanical Hamiltonian describing the CCJJ rf SQUID, as given by Eqs.~(5a)$\rightarrow$(5d) in Ref.~\onlinecite{CCJJ}, are shown as solid curves in Fig.~\ref{fig:QubitParameters}.  The model results were obtained using the means of the independently calibrated CCJJ rf SQUID parameters, namely the body inductance $L_{\text{body}}=339\pm 3\,$pH, CCJJ loop inductance $L_{\text{ccjj}}=26\pm 1\,$pH, critical current $I_q^c=3.355\pm 0.040\,\mu$A and capacitance $C_q=182\pm 4\,$fF.  Given that the experimental data agree with a multilevel quantum mechanical of the CCJJ rf SQUID whose input parameters were independently calibrated, we are confident in our identification of these devices as flux qubits \cite{CCJJ}.

The interqubit couplers and $\iqp$ compensators on this chip were calibrated according to the methods described in Ref.~\onlinecite{CJJCoupler}.  The experimental data are summarized in Fig.~\ref{fig:CouplerAndIPC}.  From the set of calibrations for all 352 interqubit couplers on the chip, we determined that the smallest maximum AFM coupling was $M_{\text{AFM}}=1.88\pm0.03\,$pH, which was readily achievable for all sixteen couplers within the unit cell used in this study.  We chose this value of $M_{\text{AFM}}$ to define the energy $J_{\text{AFM}}$ introduced in Sect.~\ref{sec:theory}.  Likewise, from the complete set of 128 $\iqp$ compensators on this chip we determined that the maximum usable AFM coupling was $\text{max}\left|M_{i}\right|=0.100\pm0.002\,$pH.  We found it convenient to use the scale factor $\alpha=2\times M_{\text{AFM}}/\text{max}\left|M_{i}\right|=37.6$ to scale the $I_q^p$ compensation signals, as discussed in Sect.~\ref{sec:theory}.

\begin{figure}[ht]
\includegraphics[width=3.25in]{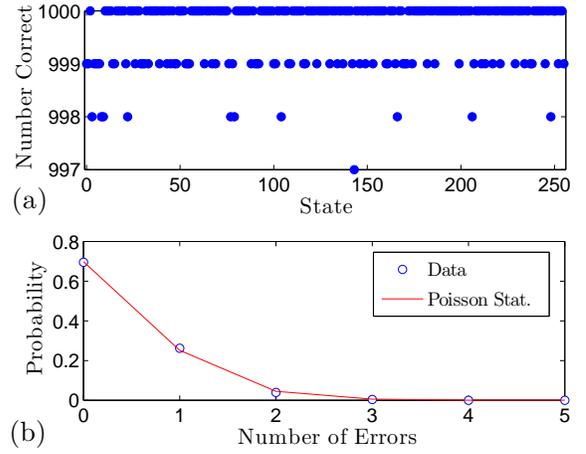}
\caption{\label{fig:Fidelity} (Color online) Verification of readout fidelity for the eight-qubit unit cell. (a) Example results from sequentially preparing the eight qubits into each of the $2^8=256$ possible final spin configurations and recording the number of correct reads for 1000 repetitions.  (b) Probability of observing a given number of errors within 1000 repetitions, as taken from the data in (a).  Data have been fit to a Poisson distribution with mean number of errors $\lambda=0.36$.}
\end{figure}

Finally, to characterize the readout fidelity of the unit cell, we set all interqubit couplers to zero coupling and then sequentially prepared and read all $2^8=256$ possible final spin configurations $\ket{\vec{s}}=\ket{s_0\;\;s_1 \ldots s_7}$, with $s_i=\pm1$.  This was accomplished by initializing the qubits in the presence of a bias vector of the form $\vec{\epsilon}\propto\left(\begin{array}{cccc} s_0 & s_1 & \ldots & s_7 \end{array}\right)$.  We prepared each configuration and read the state of the unit cell 1000 times using two repeated reads of each dc SQUID \cite{QFP}.  The results of one iteration of this fidelity check are shown in Fig.~\ref{fig:Fidelity}(a).  Here, we have adopted a shorthand notation by converting each binary word $\ket{\vec{s}}$ into a decimal number $\ket{\text{State}}$, where
\begin{displaymath}
\text{State} = \sum^7_{i=0}\frac{s_i+1}{2}2^{7-i} \; .
\end{displaymath}

\noindent In the majority of instances the readout yielded the expected result 1000 times.  We never observed more than 3 errors per 1000 repetitions.  Multiple iterations of this check revealed that results for those states that appear to be less than perfect in Fig.~\ref{fig:Fidelity}(a), such as State=$143$, were not reproducible.  Further analysis indicated that these infrequent errors were not due to problems in preparing $\ket{\vec{s}}$.  Rather, the errors were most likely generated by the statistical nature of the dc SQUID switching times \cite{QFP}.  The probability of observing a given number of errors within 1000 repetitions, as obtained by taking a histogram of the data in Fig.~\ref{fig:Fidelity}(a), is shown in Fig.~\ref{fig:Fidelity}(b).  These results are well described by Poisson statistics with a mean number of errors $\lambda=0.36$ in 1000 repetitions.  We defined the threshold for a number of counts to be statistically significant as three times the square root of the variance, $3\sqrt{\lambda}\approx 2$.  One can then deem the fidelity of our readout of the eight-qubit unit cell in these particular experiments to have been $1-3\sqrt{\lambda}/1000=0.998$.

\section{Experimental Processor Performance}
\label{sec:experiment}

To test the performance of the eight-qubit unit cell, we generated a set of 800 unique random ISG problem instances whose $h_i$ and $K_{ij}$ were drawn from a distribution of 15 evenly spaced values within the set
\begin{equation}
\label{eqn:hetK}
h_i,K_{ij}\in\left[-1\;\; -6/7 \;\; \ldots +6/7 \;\; +1\right] \;.
\end{equation}

\noindent We then took advantage of the symmetry of the physical layout of the qubits and couplers shown in Fig.~\ref{fig:DeviceSchematic}b and generated 8 permutations of each of the 800 progenitor problems by rotating and reflecting their $h_i$ and $K_{ij}$ embeddings about the center of the unit cell.  The result was a test set of 6400 problem instances, each requiring a unique embedding, that were then posed to the hardware.  Note that while other researchers have chosen to focus upon particular classes of Ising spin problems, such as random 3-SAT\cite{3-SAT}, MAX-clique\cite{MAX-CLIQUE}, and exact cover \cite{Farhi1,EXACT-COVER2}, we have found our random ISG problem set to be of great use as it encompasses a broad range of problem classes that could be posed to an AQO processor in practice.

The PMM programming required to set up the processor was rather involved and the details of its implementation are beyond the scope of this article \cite{PMM}.  Nonetheless, it is worth recognizing that, for each problem instance, the interqubit coupler biases $\Phi^x_{\text{co},ij}$ and $\iqp$-compensator biases $\Phi^x_{I_p,i}$ were reset to provide the desired $K_{ij}$ and $h_i$, respectively.  As a consequence of the changing interqubit coupler susceptibility, the $L$-tuner biases also had to be updated to hold the qubit inductance constant, as described in Ref.~\onlinecite{CCJJ}.  Finally, the effects of junction asymmetry in the CJJ interqubit couplers and $\iqp$ compensators led to a small offset of each qubit's degeneracy point $\delta\Phi^0_i$, as shown in Ref.~\onlinecite{CJJCoupler}.  These offsets were independently calibrated for each qubit prior to running the processor.  To compensate, the PMM controlling the qubit flux offset $\Phi_i^0$ was updated.

\begin{figure}[ht]
\includegraphics[width=3.25in]{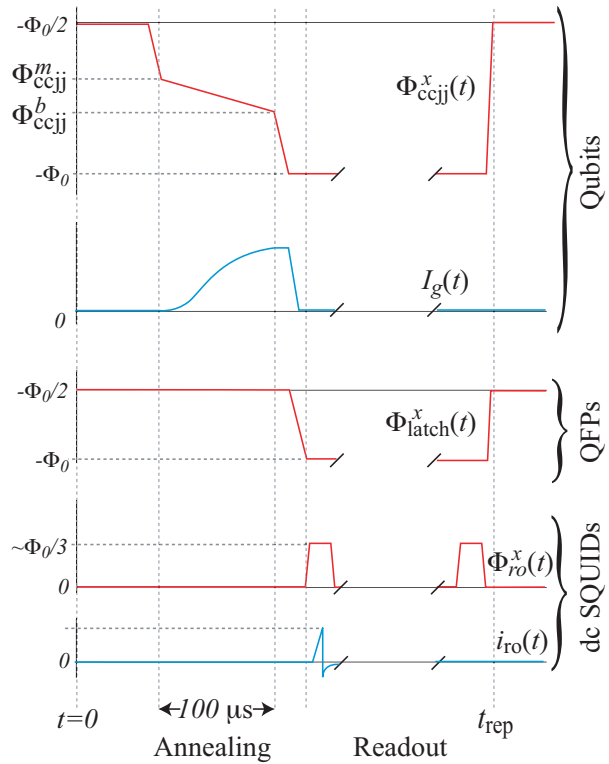}
\caption{\label{fig:Waveforms}  (Color online) Depiction of the measurement wave form sequence.  Only time-dependent biases have been shown.  All qubits are subjected to common $\Phiccjjx(t)$ and $I_g(t)$.  All QFPs are subjected to a common $\Philatchx(t)$.  Example dc-SQUID flux $\Phirox$ and current $i_{\text{ro}}(t)$ wave forms shown only for device connected to $q_0$.}
\end{figure}

With the details concerning the PMM aside, the analog control signal sequence for implementing the AQO algorithm introduced in Sect.~\ref{sec:theory} was quite simple, as depicted in Fig.~\ref{fig:Waveforms}.  During the annealing phase there were only two time-dependent signals applied to the entire unit cell: the CCJJ control signal $\Phiccjjx(t)$ and the $\iqp$-compensation signal $I_g(t)$.  For simplicity, we chose to use a relatively slow $100$-$\mu$s-long linear ramp of $\Phiccjjx(t)$ through the region relevant for qubit operation between $\Phi^m_{\text{ccjj}}/\Phi_0=-0.590$ and $\Phi^b_{\text{ccjj}}/\Phi_0=-0.650$, where the superscripts $m$ and $b$ denote biases for which the CCJJ rf-SQUID potential energy is deep in the monostable and bistable regime, respectively.  The aforementioned ramp time was sufficiently long to avoid the loss of ground state probability to Landau-Zener transitions at anticrossings \cite{IPHTLZ,OliverLandauZener,LZ}.  Experimental data and numerical simulations supporting this latter statement will be discussed in Sect.~\ref{sec:simulation}.  The time-dependent wave form $I_g(t)=\alpha\left|I^p_q(t)\right|$, with $\alpha=37.6$, was constructed by sampling the modeled $\left|I_q^p\left(\Phiccjjx(t)\right)\right|$, shown as the solid curve in Fig.~\ref{fig:QubitParameters}(a), at 200 points between $\Phi^m_{\text{ccjj}}$ and $\Phi^b_{\text{ccjj}}$.

At the end of the annealing phase, the final task that involved time-dependent wave forms was that of reading out the state of the unit cell.  The states of all eight qubits were simultaneously latched by raising the QFP tunnel barriers in unison using their global bias $\Philatchx(t)$ \cite{QFP,CCJJ}.  Thereafter, the dc-SQUID current biases and flux biases were modulated to serially read out the state of each QFP.  In Fig.~\ref{fig:Waveforms}, only those current and flux bias wave forms for operating the readout connected to $q_0$ have been depicted.  The entire wave form sequence, including programming of the 992 PMM elements, a $1\,$ms cooling period following PMM programming, and 1000 repetitions of the anneal and read sequence depicted in Fig.~\ref{fig:Waveforms} was completed in $\approx2.5\,$s.

The key experimental parameter that was varied in this study was the temperature of the processor $T$.  We calibrated $T$ in situ using macroscopic resonant tunneling (MRT) measurements versus $\Phiqx$ at $\Phiccjjx/\Phi_0=-0.6238$ using a dwell time of $1\,$ms on qubits $q_0$ and $q_4$ \cite{MRT}.  At this particular CCJJ bias, the qubits reached thermal equilibrium within the specified dwell time and the resultant population statistics versus $\Phiqx$ could be fit to the form
\begin{displaymath}
P_{\text{MRT}}=\frac{1}{2}\left[1+\tanh\left(\frac{2\iqp\left(\Phiqx-\Phi_q^0\right)}{2k_BT}\right)\right] \; ,
\end{displaymath}

\noindent where $\iqp=1.26\pm 0.02\,\mu$A was obtained from the independent calibration shown in Fig.~\ref{fig:QubitParameters}(a) and $\Phi_q^0$ was the independently calibrated qubit flux offset.  This then left only one free parameter, namely $T$.

\begin{figure}[ht]
\includegraphics[width=3.25in]{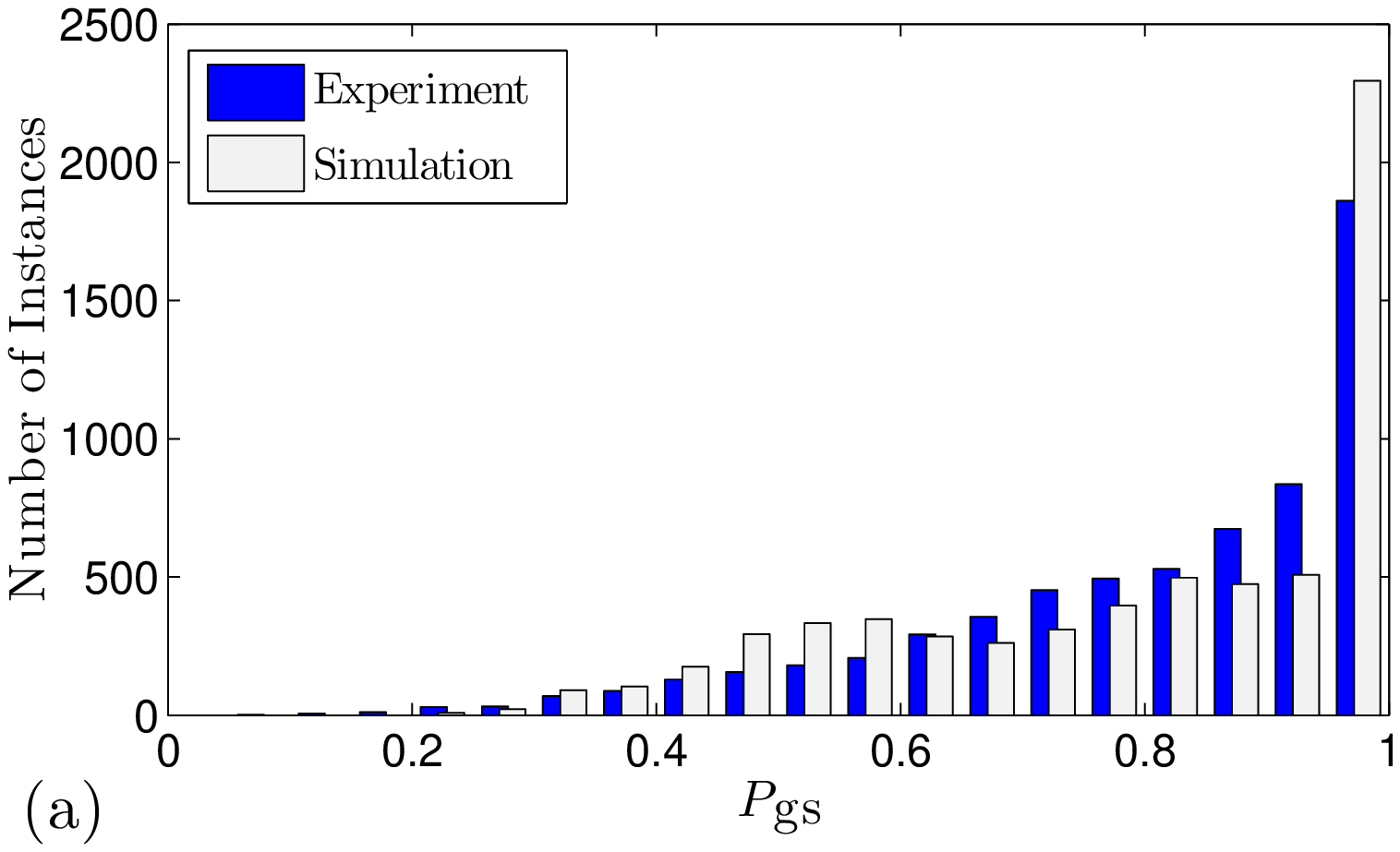} \\
\includegraphics[width=3.25in]{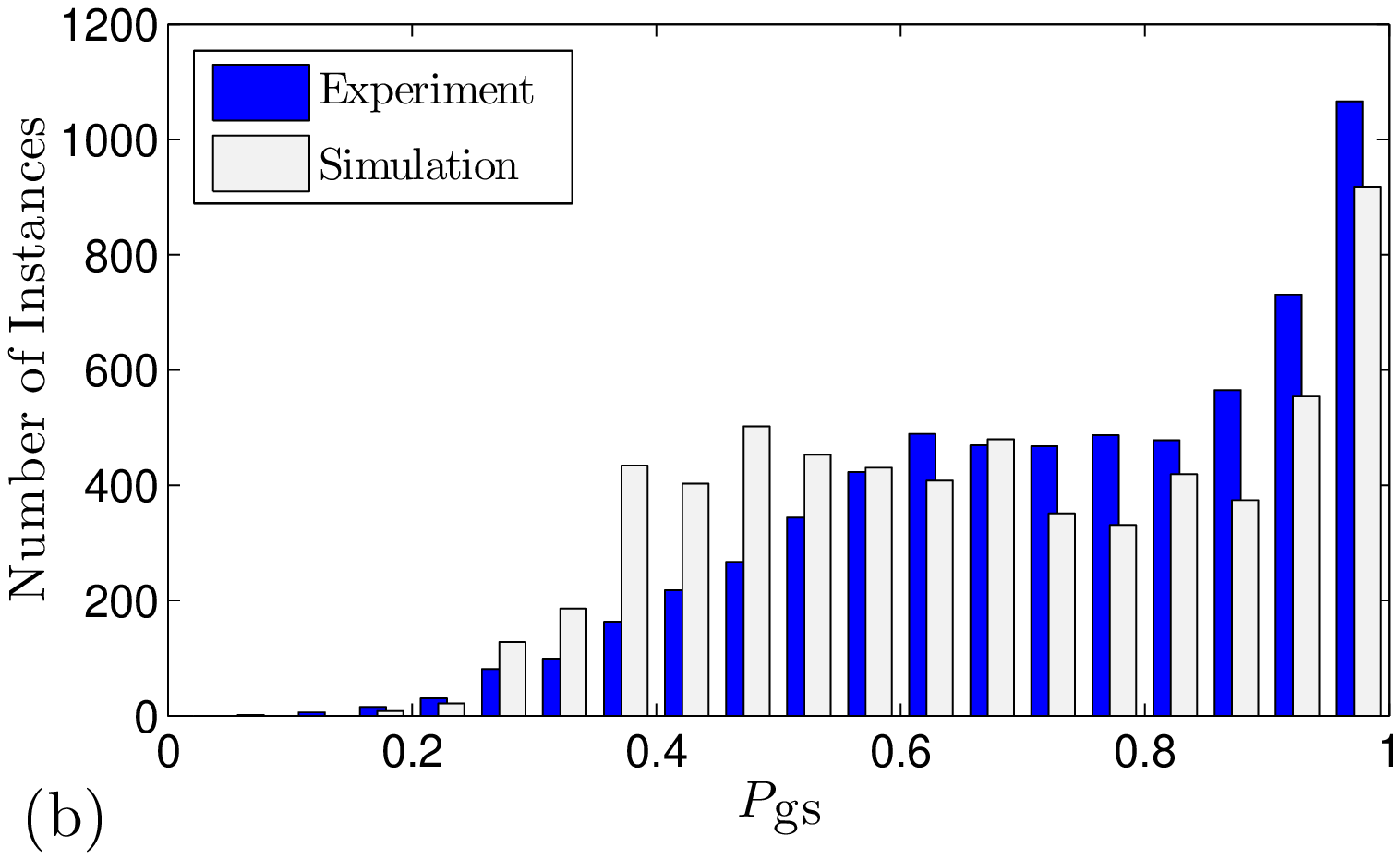} \\
\includegraphics[width=3.25in]{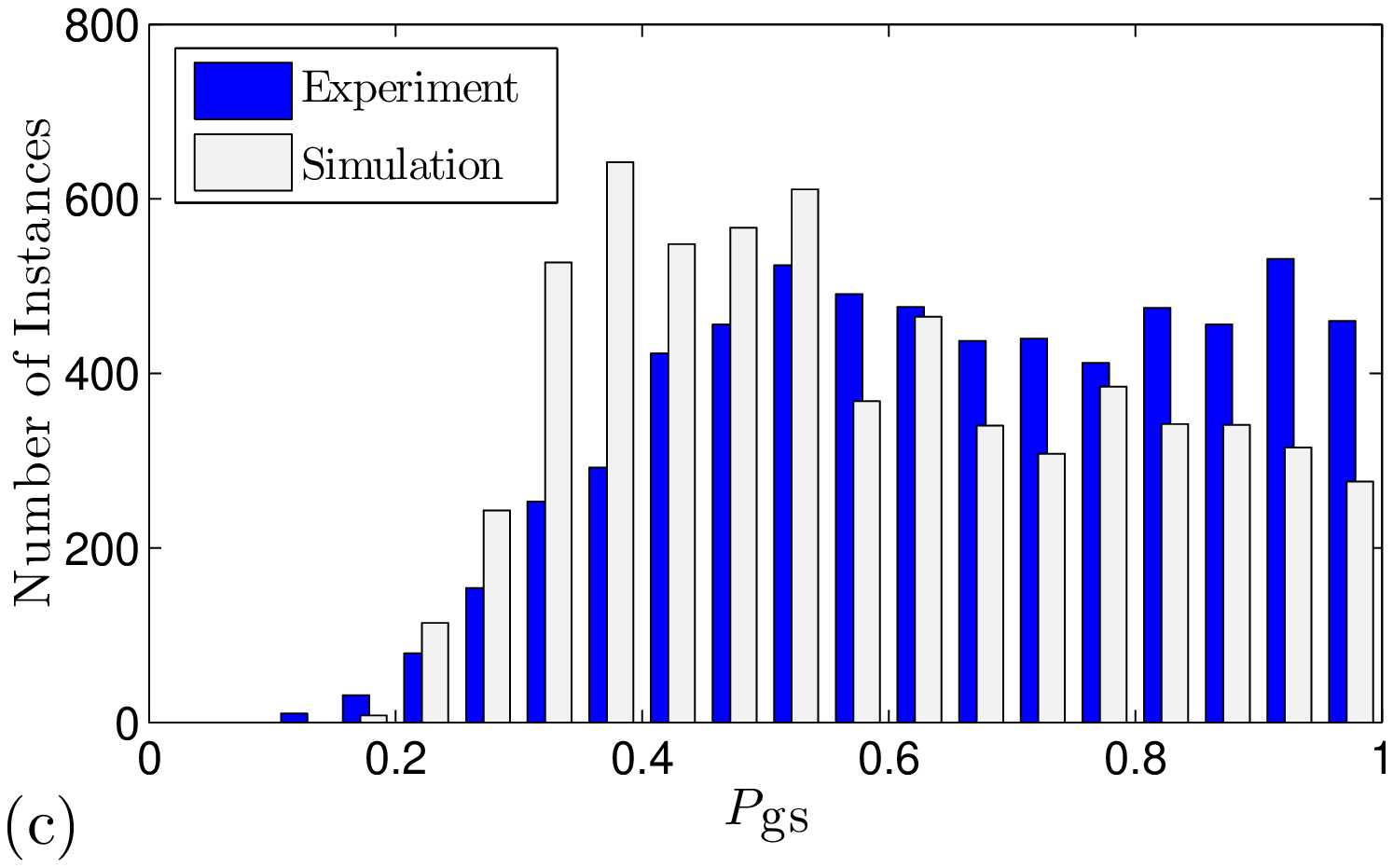}
\caption{\label{fig:PCorrect} (Color online) Histograms of the experimentally observed and the simulated probability of finding the system in the ground state $P_{\text{gs}}$ from a test involving 6400 unique random ISG instances at three different temperatures: (a) $T=20\,$mK, (b) $T=35\,$mK, and (c) $T=50\,$mK.}
\end{figure}

The unit cell was tested using the set of 6400 problem instances at three values of $T$.  Each instance was run 1000 times and the probabilities of observing each of the 256 possible $\ket{\vec{s}}$ were recorded.  Given the small size of the problems it was trivial to independently determine $\ket{\vec{s}_{\text{gs}}}$.  Histograms of the probability of observing the ground state $P_{\text{gs}}$ are shown in Fig.~\ref{fig:PCorrect}.  The output of numerical simulations shown in these plots will be discussed in Sect.~\ref{sec:simulation}.  Here, the vertical axes correspond to the number of problem instances for which the hardware returned $P_{\text{gs}}$ within the bounds of each bin of the histogram.  The data taken at $T=20\,$mK, see Fig.~\ref{fig:PCorrect}(a), show an obvious spike in the highest bin, $0.95<P_{\text{gs}}\leq1$, and a tail that appears to be exponentially suppressed at lower $P_{\text{gs}}$.  For $T=35\,$mK, see Fig.~\ref{fig:PCorrect}(b), the spike in the highest bin appears diminished and a broad hump in the vicinity of $P_{\text{gs}}\sim 0.6$ is observed.  Finally, for $T=50\,$mK, see Fig.~\ref{fig:PCorrect}(c), the high $P_{\text{gs}}$ spike has been flattened and the broad hump has shifted to slightly lower $P_{\text{gs}}\sim 0.5$.  Nonetheless, for all three values of $T$, $P_{\text{gs}}>0.1$ for all instances.  

\begin{figure}[ht]
\includegraphics[width=3.15in]{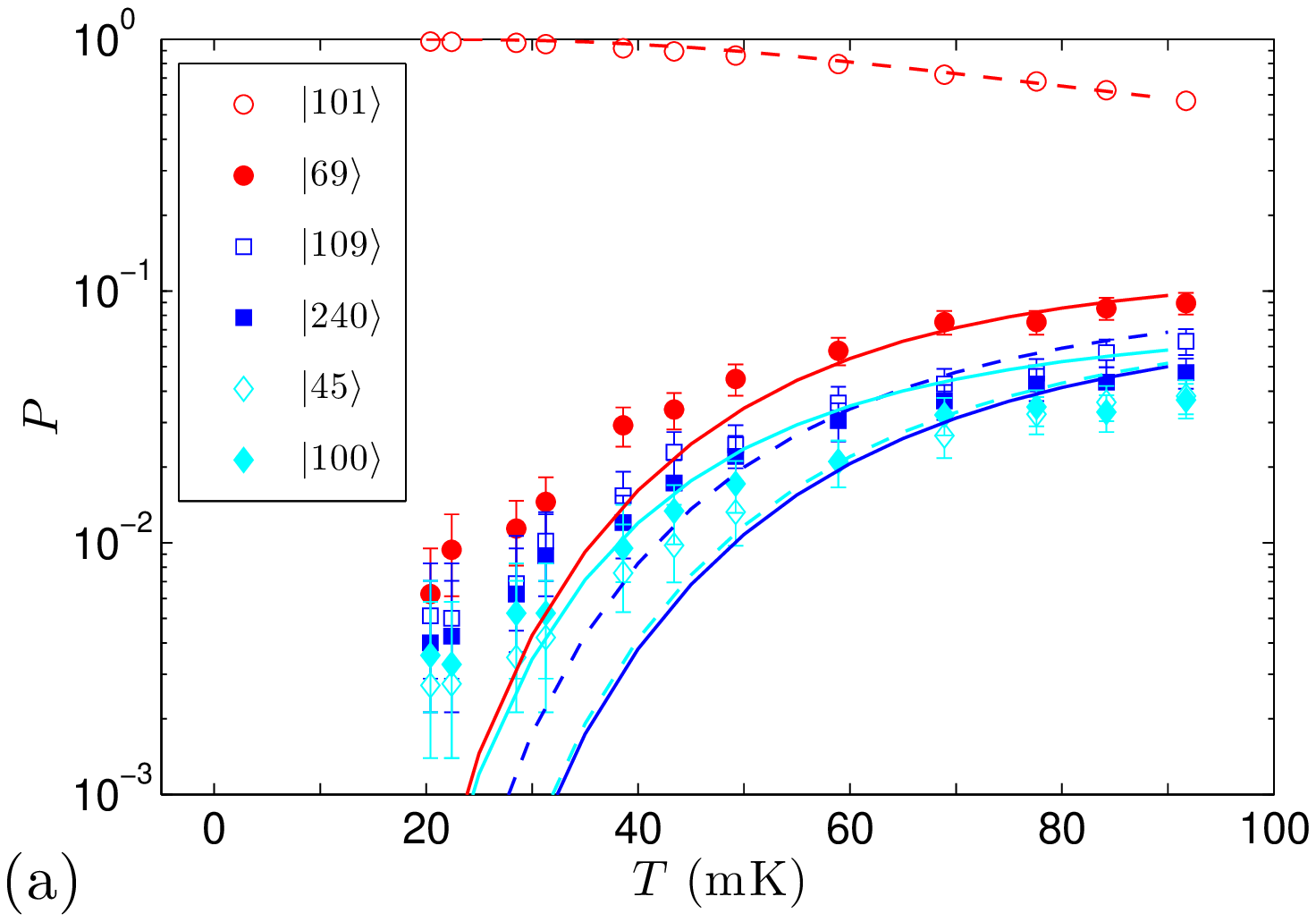} \\
\includegraphics[width=3.15in]{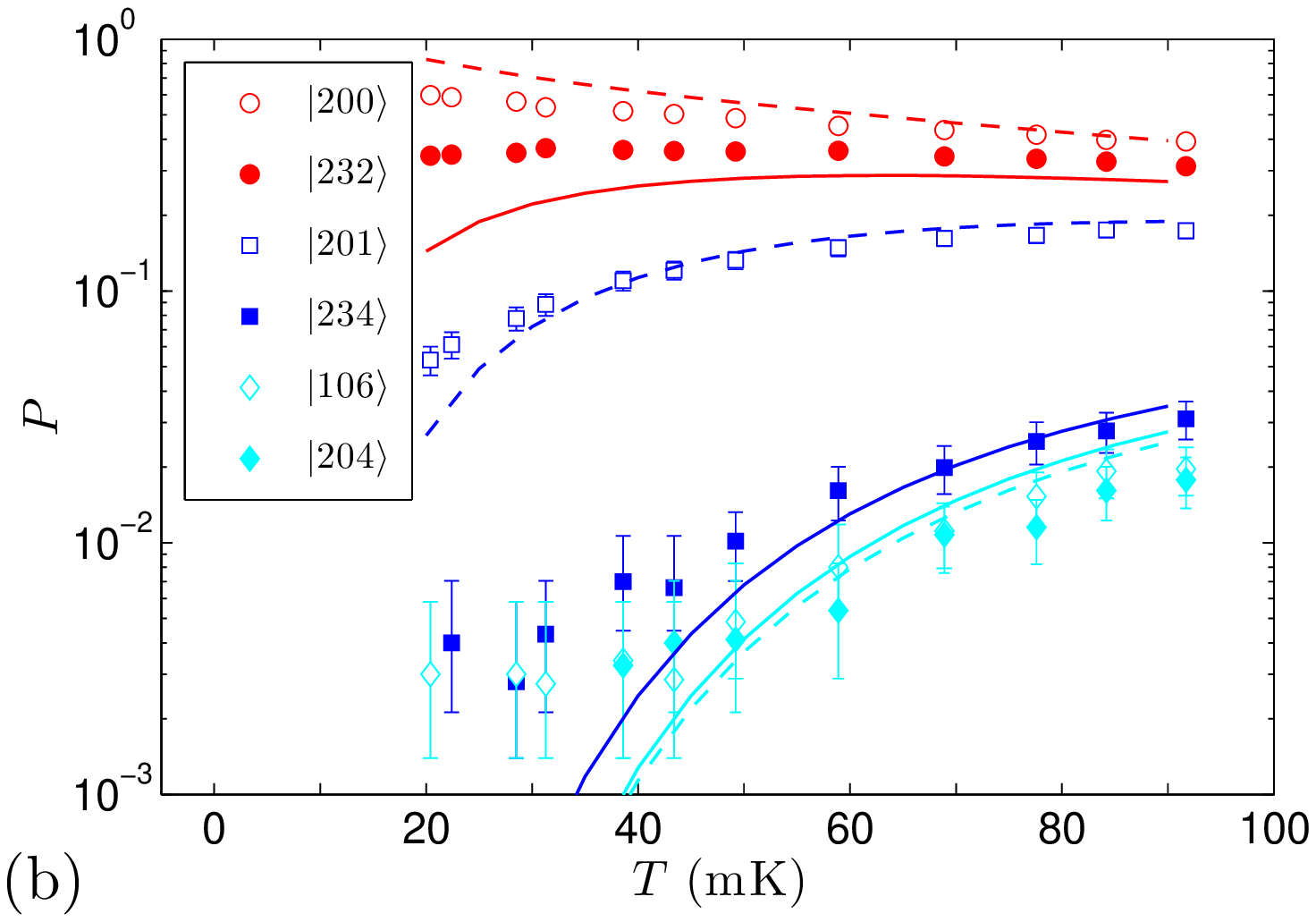} \\
\includegraphics[width=3.15in]{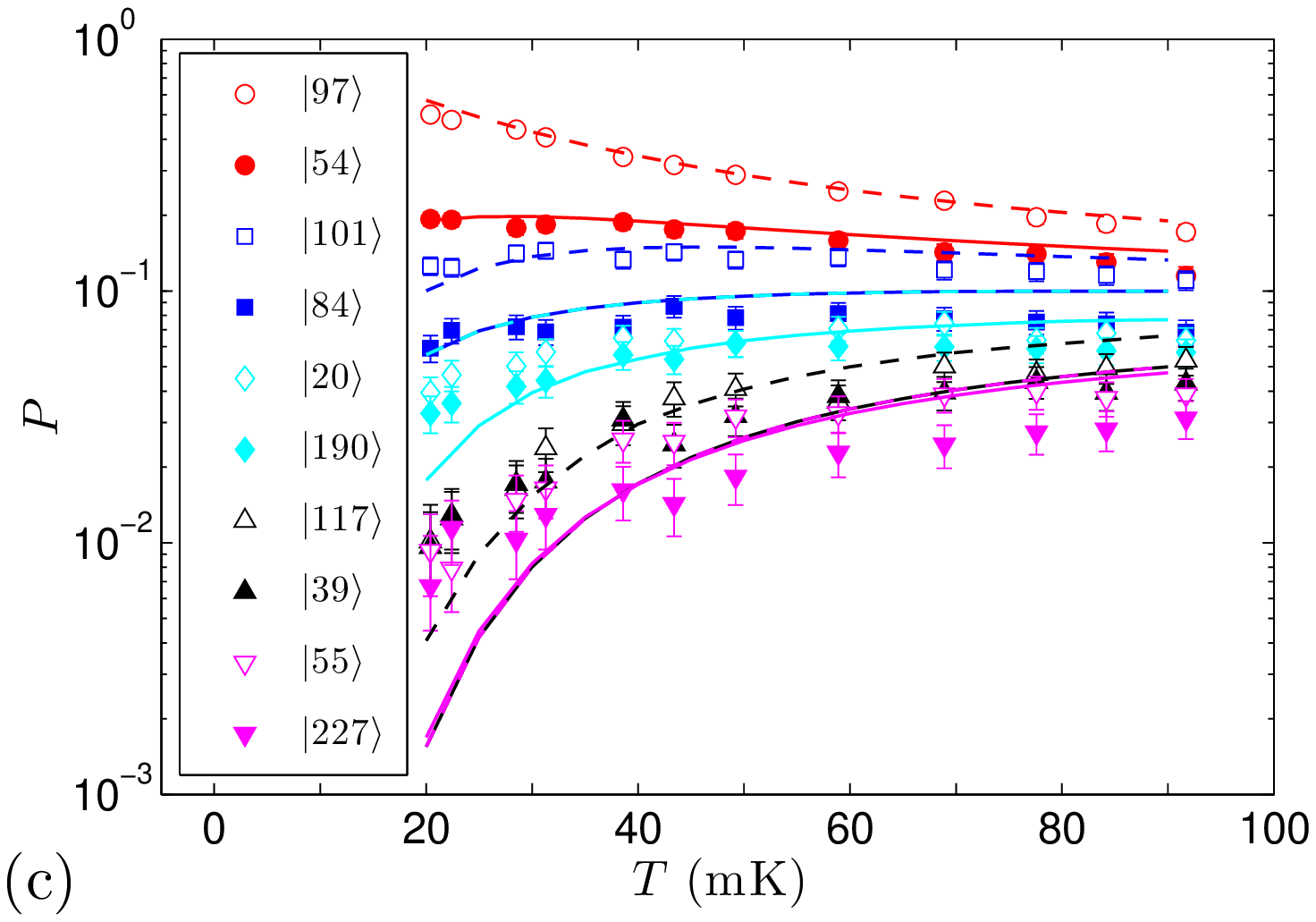}
\caption{\label{fig:TSweep} (Color online) Experimental (symbols) and simulated (curves) probabilities of finding the system in low energy states. (a) Example A. (b) Example B. (c) Example C.  Solid (dashed) curves have been color-coded to match filled (hollow) symbols.}
\end{figure}

To gain further insight into the source of the structure of the histograms shown in Fig.~\ref{fig:PCorrect}, we selected an exemplary problem instance from each of the three features in the $T=35\,$mK results: Example A from the spike ($P_{\text{gs}}\sim 1$), Example B from the broad hump ($P_{\text{gs}}\sim 0.6$), and Example C from the tail ($P_{\text{gs}}\sim 0.4$).  The instance settings $h_i$ and $K_{ij}$ for these three examples are given in Appendix \ref{InstanceSettings}.  We then ran each of these example instances 1000 times at a series of temperatures $20\,\text{mK}\leq T<100\,\text{mK}$.  A summary of the observed population statistics $P$ is shown in Fig.~\ref{fig:TSweep}.   The output of numerical simulations shown in these plots will be discussed in Sect.~\ref{sec:simulation}.  For Examples A, B, and C, a total of 28, 13, and 38 states were observed above the readout resolution threshold at $T=91.7\,$mK, respectively.  We have plotted the data for only those states that were observed with $P>0.01$ at $T=91.7\,$mK.  

\begin{figure}[hb]
\includegraphics[width=3.15in]{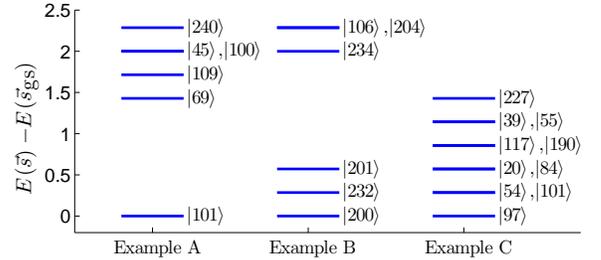}
\caption{\label{fig:ExampleObjective} (Color online) Evaluation of the dimensionless objective function $E(\vec{s})$ [Eq.~(\ref{eqn:QUBO})] for Examples A, B, and C.  Instance settings are given in Appendix ~\ref{InstanceSettings}.  Results plotted relative to the lowest value $E(\vec{s}_{\text{gs}})$.  Spin configurations $\ket{\vec{s}}$ are identified to the right of each level.}
\end{figure}

The results shown in Fig.~\ref{fig:TSweep} indicate that a small number of $\ket{\vec{s}}$ are returned for any given problem instance and that the probability of observing those states evolves smoothly with $T$.  To assist in determining why particular states were observed, we have evaluated the objective function $E(\vec{s})$, given by Eq.~(\ref{eqn:QUBO}), for Examples A, B, and C using the values of $h_i$ and $K_{ij}$ provided in Appendix \ref{InstanceSettings}.  The results 
have been summarized in Fig.~\ref{fig:ExampleObjective}.  This plot reveals that the ground state $\ket{\vec{s}_{\text{gs}}}=\ket{101}$, $\ket{200}$, and $\ket{97}$ for Example A, B, and C, respectively.  Likewise, the first excited state(s) can be identified as $\ket{\vec{s}}=\ket{69}$, $\ket{232}$, and the degenerate pair ($\ket{54}$,$\ket{101}$) for Example A, B, and C, respectively.  From a comparison of Figs.~\ref{fig:TSweep} and \ref{fig:ExampleObjective} one can make two important observations:  First, for all three examples the order of the observed states, from most to least probable, {\it exactly} matches the order when ranked by ascending $E(\vec{s})$.  Thus the ground state is the most probable state, the first excited state is the second most probable, and so forth.  The only exception is the order of $\ket{240}$ and ($\ket{45}$,$\ket{100}$) in Example A, though this discrepancy is within the noise.  Second, there is a preferred order for some of the pairs of degenerate excited states such as ($\ket{54}$,$\ket{101}$) and ($\ket{190}$,$\ket{117}$) for Example C.  While the preference for one over the other appears to be subtle, it is nonetheless very reproducible.
 
To summarize this section of the paper, we have tested an eight-qubit unit cell inside a superconducting processor that embodies an AQO algorithm.  The unit cell has been tested by running 6400 random ISG problem instances at three temperatures and three representative instances at multiple temperatures.  With repeated measurements, the hardware returns a small number of final spin configurations, with the ground state always being the most probable result.  The remaining states that are observed 
correspond to low lying excited states.

\section{Simulated Processor Performance}
\label{sec:simulation}

The objective of this section of the paper is to present a physical picture that naturally explains the final distribution of $\ket{\vec{s}}$ that was obtained from any given ISG problem that was posed to the hardware.  This picture relies upon having independently calibrated all qubit, coupler, and $\iqp$-compensator parameters that enter into Hamiltonian (\ref{eqn:Hrfsrearranged}) and then adding interactions between all eight qubits and an environment.  It will be demonstrated that the inclusion of quantum mechanical relaxation \cite{Weiss} is sufficient to explain the observed distributions.

We have chosen to model this system by neglecting all but the two lowest lying states in each CCJJ rf SQUID (qubit approximation).  Furthermore, we have also assumed that the qubit parameters $\left|I_q^p(\Phiccjjx)\right|$ and $\Delta_q(\Phiccjjx)$ are the same for all qubits and are given by the solid curves in Fig.~\ref{fig:QubitParameters}.  This latter assumption would not be justified without having first leveraged all of the features of the CCJJ rf SQUID so as to synchronize their qubit parameters to a reasonable level.  The low energy physics of the ideal closed system will then be dictated by Hamiltonian (\ref{eqn:Hrfsrearranged}).    

\begin{figure}[ht]
\includegraphics[width=3.25in]{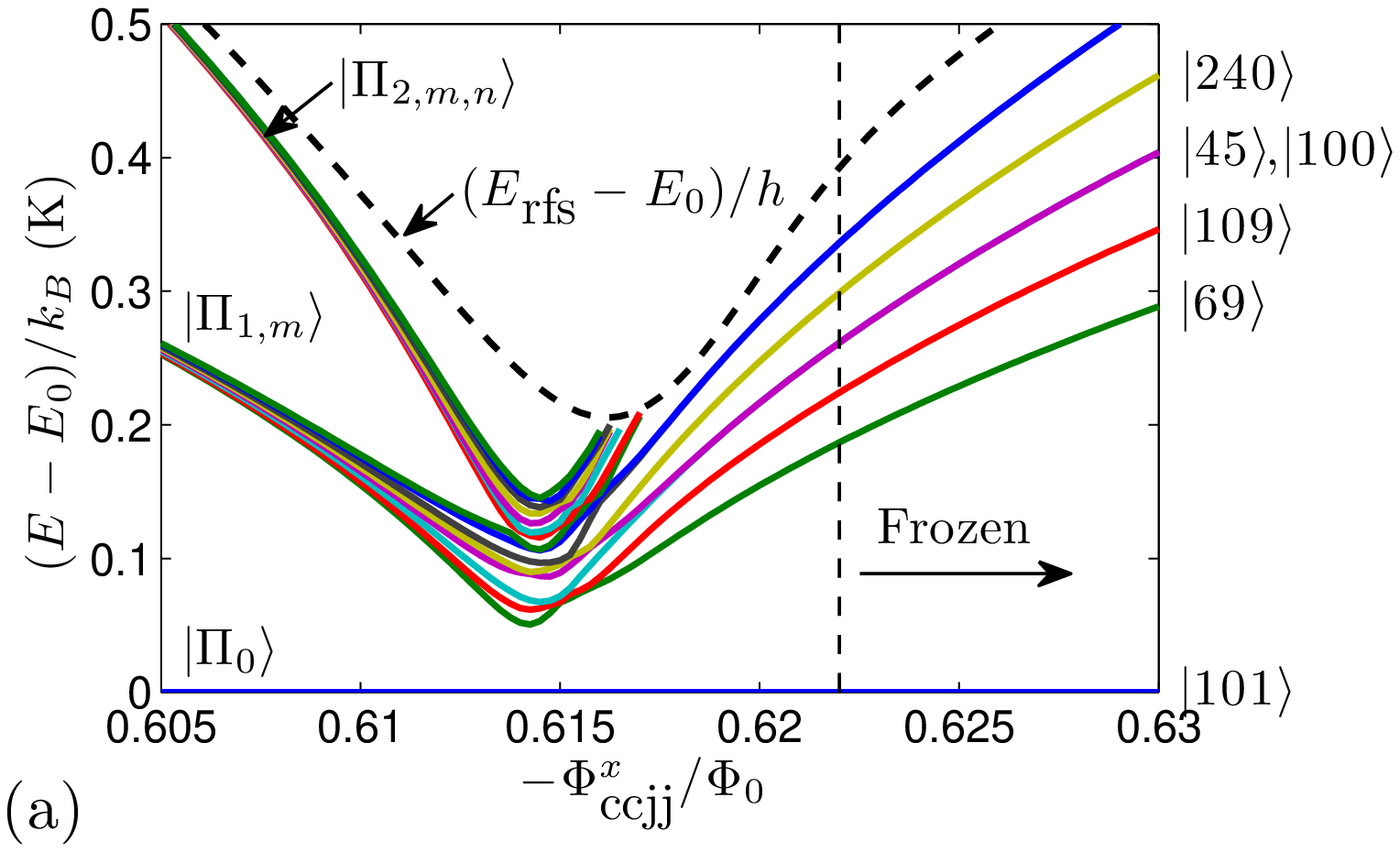} \\
\includegraphics[width=3.25in]{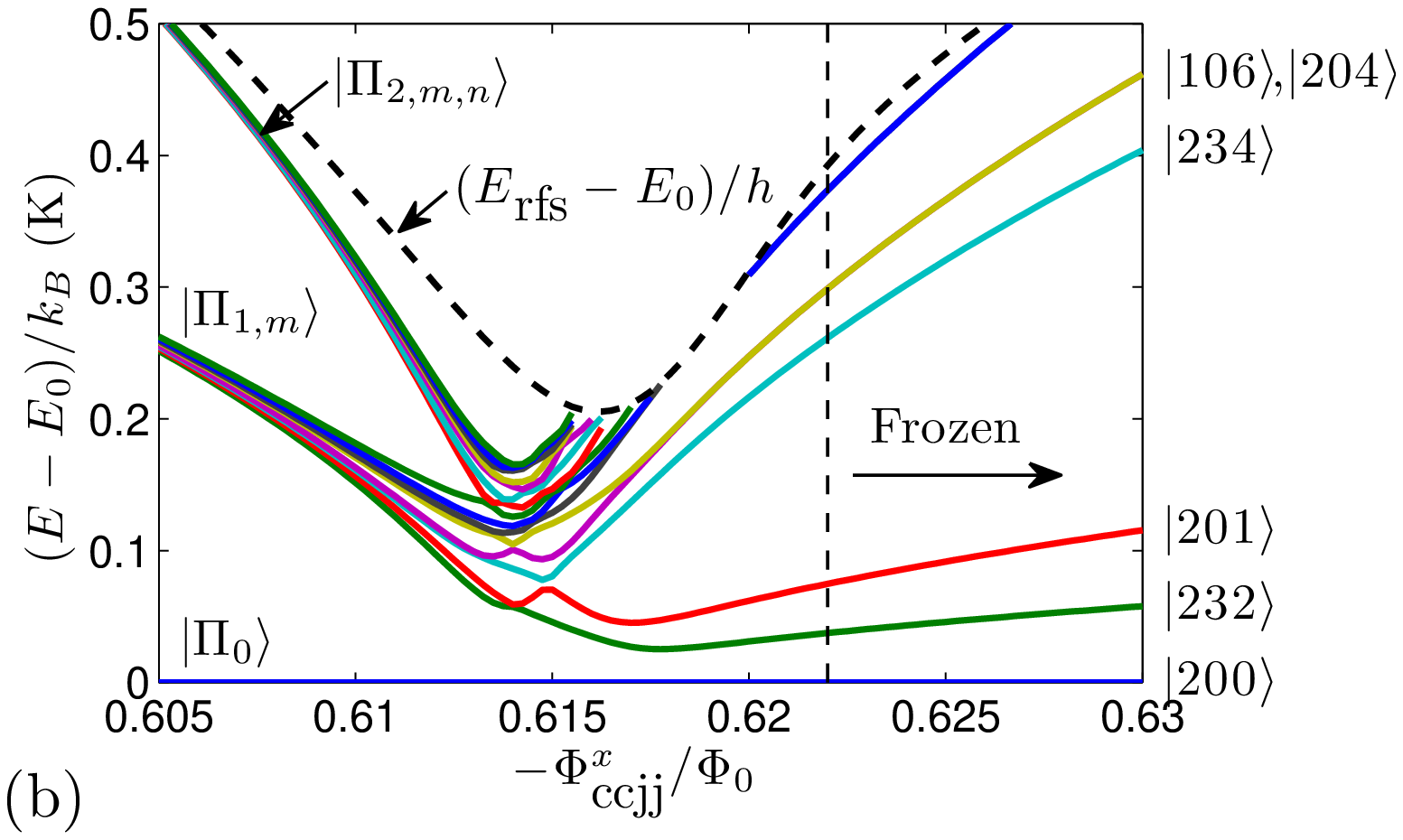} \\
\includegraphics[width=3.25in]{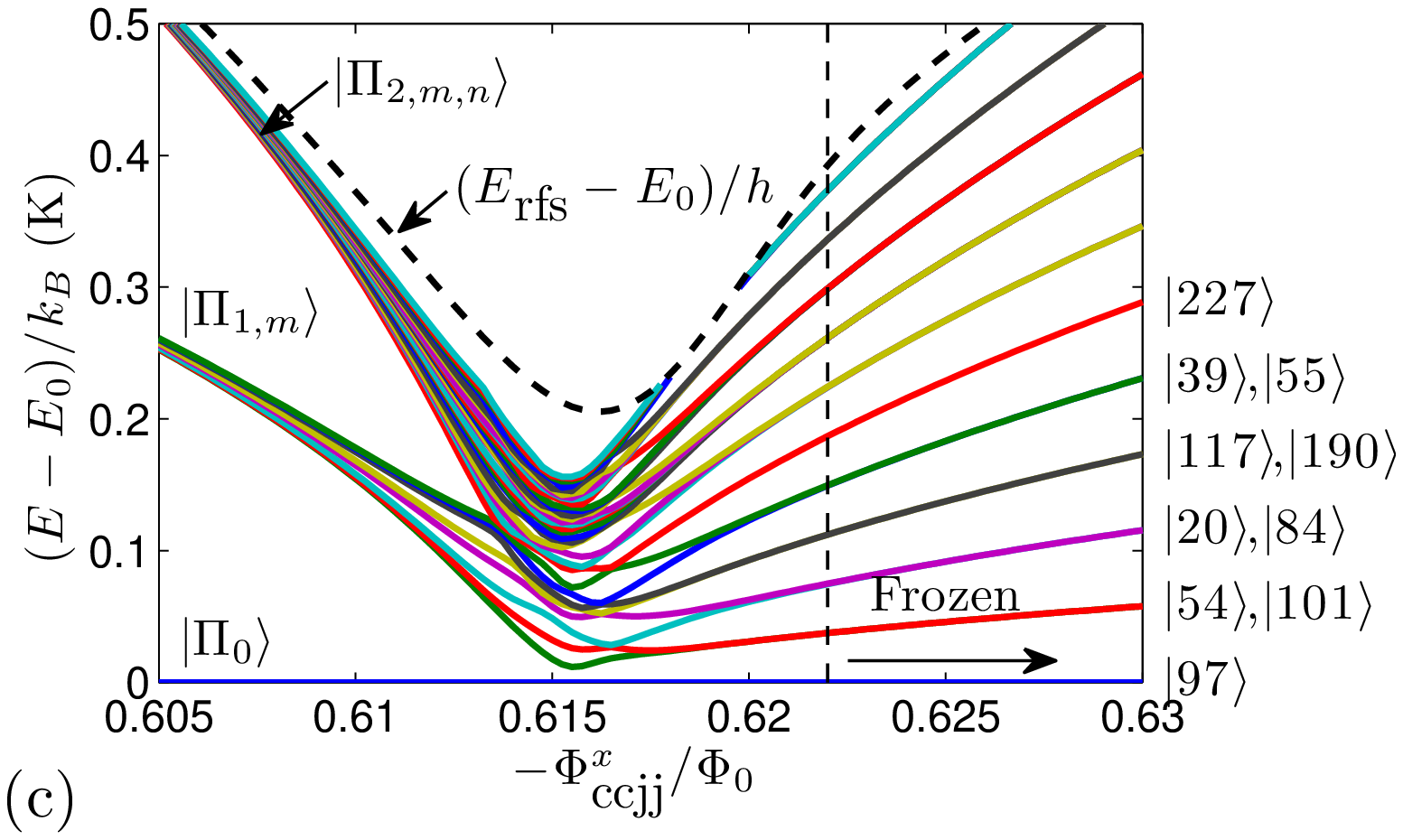}
\caption{\label{fig:Eigenspectra} (Color online) Calculated low energy eigenspectra versus $-\Phiccjjx$.  (a) Example A. (b) Example B.  (c) Example C.  Energies have been plotted relative to the ground state energy $E_0$.  Energies have been truncated above that of the second excited state of the CCJJ rf SQUID, $E_{\text{rfs}}-E_0$.  Initial states are labeled as $\ket{\Pi_0}$, $\ket{\Pi_{1,m}}$ and $\ket{\Pi_{2,m,n}}$ on the left.  Final states are labeled as $\ket{\vec{s}}$ on the right.  Vertical dashed lines roughly delineate the value of $-\Phiccjjx$ beyond which the population statistics appear to freeze during annealing.}
\end{figure}

Instantaneous eigenspectra as a function of annealing parameter $\Phiccjjx$ for example problem instances A, B, and C, as generated using Hamiltonian (\ref{eqn:Hrfsrearranged}), are shown in Fig.~\ref{fig:Eigenspectra}.  Note the negative sign in front of the independent variable in these plots, which has been inserted such that annealing progresses from left to right.  The eigenenergies $E$ have been plotted relative to the ground state energy $E_0$ such that the ground state is coincident with the horizontal axis.  The lowest 16 levels have been plotted for Examples A and B and the lowest 32 levels for Example C.  In all cases the eigenspectra have been truncated above the second excited state of an isolated CCJJ rf SQUID, here denoted as $E_{\text{rfs}}$, as Hamiltonian (\ref{eqn:Hrfsrearranged}) is not the correct representation of the physics of the closed system beyond this energy.  This latter point is of significant practical interest and will receive more attention in Sect.~\ref{sec:discussion}.  The initial states have been identified on  the left of each plot.  Here, the ground state is denoted as 
\begin{displaymath}
\ket{\Pi_0}=\frac{1}{16}\prod^7_{i=0}\left(\ket{-1}_i+\ket{+1}_i\right)
\end{displaymath}

\noindent and is a product of the isolated qubit ground states.  There is an octet of first excited states, denoted as $\ket{\Pi_{1,m}}$, which are product states of all qubits in their ground state except a single qubit $m$ in its excited state, $\left(\ket{-1}_m-\ket{+1}_m\right)/\sqrt{2}$.  Analogously, there are 64 second excited states, denoted as $\ket{\Pi_{2,m,n}}$, in which all but two qubits are found in their ground state.  In all cases, the excited states drop in energy with increasing annealing parameter until they reach a minimum in the vicinity of $\Phiccjjx/\Phi_0\sim -0.615$.  Each of the minima correspond to a point in the annealing process at which the ground state, initially $\ket{\Pi_0}$, sheds a localized spin configuration $\ket{\vec{s}}$ from the superposition, thus eventually localizing in the solution state $\ket{\vec{s}_{\text{gs}}}$.  Likewise, the excited states become localized in configurations $\ket{\vec{s}}\neq\ket{\vec{s}_{\text{gs}}}$ as they are ejected from the ground state, and then increase in energy to the right of the minima.  In a large-scale QSG, this process would give rise to a second-order phase transition between a ground state possessing a high degree of symmetry (a superposition of all $\ket{\vec{s}}$) to the left of the minima and a ground state with broken symmetry (localized in $\ket{\vec{s}_{\text{gs}}}$) to the right of the minima.   Those $\ket{\vec{s}}$ that were observed with $P>0.01$ at $T=91.7\,$mK, as indicated in Fig.~\ref{fig:TSweep}, have been labeled on the right of each corresponding plot in Fig.~\ref{fig:Eigenspectra}.

Each plot in Fig.~\ref{fig:Eigenspectra} has been marked with a vertical dashed line at $\Phiccjjx/\Phi_0= -0.622$ to roughly delineate where in the annealing process the population statistics appear to freeze.  As a crude approximation, one can take the energies of the excited states along this vertical line and calculate Boltzmann occupation factors that roughly agree with the results presented in Fig.~\ref{fig:TSweep}.  However, such a simplistic picture cannot explain the very reproducible disparity between the observed probabilities of excited states that are degenerate along the vertical line.

To begin the discussion of our dynamical model, it is prudent to briefly address what is {\it not} the source of either the structure of the $P_{\text{gs}}$ histograms in Fig.~\ref{fig:PCorrect} or the nonzero probability of observing excited states in Fig.~\ref{fig:TSweep}.  We argue that none of these observations can be explained by a violation of the adiabatic condition.  First, we have studied all 800 progenitor random ISG problems by numerically solving a dynamical model the system described by Hamiltonian (\ref{eqn:Hrfsrearranged}) in the absence of environmental noise and subject to the wave forms depicted in Fig.~\ref{fig:Waveforms}.  These simulations revealed that the $100\,\mu$s CCJJ bias ramp time was sufficiently long as to deem the evolution of the closed system adiabatic, as they always returned $P_{\text{gs}}=1$ to within numerical error.  Second, we have experimentally studied Examples A, B, and C at multiple CCJJ bias ramp times from $10\,\mu$s to $1\,$ms and observed relatively little change in the outcome.  Bandwidth limits on our apparatus prevented us from studying the system subject to shorter ramp times.

We now turn to a description of the open system model that we have investigated.  We wish to first deal with the physics of relaxation and then discuss the effects of dephasing thereafter.  Furthermore, we must carefully distinguish between AQO, in which the evolution of the closed system can be described as being adiabatic, and QA, in which the coupled qubit system exchanges energy with its environment.

It has been argued that flux noise is the principal source of decoherence for superconducting qubits of the type included in our circuits \cite{MRT}.  For solid state qubits, where materials defects proximate to the qubit body are most likely the intermediaries by which a thermal environment disturbs the state of a qubit, this is best captured by coupling each qubit to its own independent local environment.  The Hamiltonian for this open system can be written as
\begin{equation}
\label{eqn:Hopen}
{\cal H}(t)= \ham(t) +\frac{1}{2}\sum^7_{i=0}2\left|I_q^p\left(t\right)\right|Q_i(t)\sigma_z^{(i)} \, ,
\end{equation}

\noindent where $\ham(t)$ is given by Hamiltonian (\ref{eqn:Hrfsrearranged}) and $Q_i(t)$ is an operator that acts upon the environment seen by qubit $i$ that has units of flux.  Note the factor of $\left|I_q^p\left(t\right)\right|$ preceding each $Q_i(t)$, which captures the fact that as annealing progresses that the coupling between qubit $i$ and environmental flux noise will change.  The noise spectral density for the environment surrounding qubit $i$ as a function of angular frequency $\omega$ can be expressed as \cite{Schoelkopf}
\begin{equation}
\label{eqn:correlator}
S_i(\omega)\equiv\left(2\left|I_q^p\left(t\right)\right|\right)^2\int dt^{\prime} e^{i\omega t^{\prime}}\left<Q_i(t^{\prime})Q_i(0)\right> \, .
\end{equation}

\noindent  For Markovian noise \cite{Weiss}, the fluctuation-dissipation theorem allows one to write
\begin{equation}
\label{eqn:S}
S_i(\omega)=\frac{2\text{Im}\left\{L_0(\omega)\right\}\left(2\left|I^p_q(t)\right|\right)^2\hbar\omega}{1-\exp\left(-\frac{\hbar\omega}{k_BT}\right)} \; ,
\end{equation}

\noindent where $L_0(\omega)$ can be thought of as an inductance per unit bandwidth that is imparted to qubit $i$ by its local environment.  At a more fundamental level, $L_0$ must be proportional to the environment's complex magnetic susceptibility summed over all modes.

While there is general consensus that the flux noise power spectral density in superconducting flux qubits is $1/f$ in nature at very low frequency \cite{1OverF}, relatively little is known about the form of $S_i(\omega)$ that is appropriate at the high energies required to induce transitions in these qubits.  In lieu of a more detailed picture of $S_i(\omega)$, we have chosen to follow a simple but physically reasonable approach by taking the Ohmic approximation in which Im$\left\{L_0(\omega)\right\}$ is treated as a constant.  This simple quantum mechanical model of an environment is sufficient for our purposes as it provides an asymmetric $S_i(\omega)$ with respect to $\omega$.  Consequently, the fluctuation-dissipation theorem ensures that there are thermally occupied modes of the environment that can fuel excitations of the eight-qubit system, thus compromising $P_{\text{gs}}$.

To model the dynamics of the open system embodied by Hamiltonian (\ref{eqn:Hopen}), we have constructed a numerical model based upon a generalized Bloch-Redfield formalism \cite{Blum,BR}.  In this model, one integrates a set of coupled first order differential equations for the elements of the eight-qubit system's density matrix $\rho(t)$, as given by
\begin{subequations}
\begin{equation}
\label{eqn:BR1}
\partial_t\rho_{mn} =  -i\omega_{mn}\rho_{mn}-\sum_{k,l}\left(R_{mnkl}+M_{mnkl}\right)\rho_{kl} \, ,
\end{equation}
\begin{equation}
\label{eqn:BR2}
M_{mnkl}=-\delta_{nk}\bra{l}\partial_t\ket{m}-\delta_{ml}\bra{n}\partial_t\ket{k} \, ,
\end{equation}
\end{subequations}

\noindent where $R_{mnkl}$ is the so-called Redfield tensor \cite{BR} which is constructed from matrix elements of the flux operators between the eight-qubit states $\ket{m}$ and the $S_i(\omega)$ of the individual qubits.  These matrix elements, which vary proportional to the number of qubit flips between a pair of states, are what facilitates thermalization of the eight-qubit system.  In order to make the numerical work manageable on a personal computer, we kept only the ten lowest lying states.  We have verified that the output of this model in the limit of small Im$\left\{L_0\right\}$ matches that of numerically solving the Schrodinger equation.

Our numerical model of the open system contains only one free parameter, namely, Im$\left\{L_0\right\}$.  One can generate a crude estimate of this parameter based upon the width $W$ of a single qubit lowest order MRT peak deep in the incoherent regime.  From Eq.~(9) of Ref.~\onlinecite{AminAverin},
\begin{equation}
\label{eqn:W}
W^2 =\int \! d\omega S_i(\omega) \; .
\end{equation}

\noindent For the qubits in the circuit studied in this paper, experimental results yielded Gaussian-shaped MRT peaks of width $W/2\iqp\sim 80\,\mu\Phi_0$ at $T=20\,$mK.  Low frequency drift measurements, of the type described in Ref.~\onlinecite{1OverFGeometry}, revealed low frequency flux noise with a one-sided power spectral density that could be fit to the form $A^2/f$ with $A\sim 1\mu\Phi_0/\sqrt{\text{Hz}}$.  Integrating this $1/f$ power spectrum over a nominal ten decades in frequency yields a root mean squared flux noise on the order of $W_{\text{LF}}/2\iqp\sim 5\,\mu\Phi_0$.  Therefore, there was an additional $\left(W-W_{\text{LF}}\right)/2\iqp\sim 75\,\mu\Phi_0$ of integrated flux noise that was not captured by the $1/f$ portion of the noise spectrum.  Assuming that the remainder of $W/2\iqp$ can be attributed to a small white flux noise level that is integrated out to a cutoff $f_c=k_BT/h\sim 460\,$MHz, we estimate that the amplitude of the one-sided flux noise spectral density will be on the order of $\delta\Phi_n\sim 3.5\,$n$\Phi_0/\sqrt{Hz}$.  This white noise level is well below the detection limits realized in all low frequency flux noise measurements on SQUIDs reported in the literature to date.  Experimental evidence of an even lower white flux noise level in a 3-junction rf SQUID flux qubit has been reported in Ref.~\onlinecite{Deppe}.  In terms of an environmental inductance, our estimate of $\delta\Phi_n$ translates into Im$\left\{L_0\right\}=47\,$yH$/$Hz.

The results of numerical simulations of Eq.~(\ref{eqn:BR1}) with the estimated Im$\left\{L_0\right\}$ and the problem instance settings for Examples A, B, and C as a function of $T$ are summarized in Fig.~\ref{fig:TSweep}.  Here, the coloring and format of the curves have been chosen to correspond to the color and fill of the data symbols in those plots, with hollow (solid) points corresponding to dashed (solid) curves.  While the match between theory and experiment is not perfect, it is nonetheless clear that the generalized Bloch-Redfield model does provide a reasonable prediction of the populations for all three Examples.  We believe that most of the discrepancies can be explained by slight errors in problem specification, as will be discussed below, and therefore did not pursue any fitting of these data by changing Im$\left\{L_0\right\}$.  Very importantly, we draw attention to the splitting of the population statistics between the pairs of degenerate excited states ($\ket{54}$,$\ket{101}$) and ($\ket{190}$,$\ket{117}$) for Example C.  These features arise quite naturally from the model and are a result of differences in the transition rates between states during the course of annealing, as governed by the quantum mechanical matrix elements of the flux operators between pairs of initial and final states.

In addition to the three examples cited above, we have simulated all 6400 problem instances using our model and constructed $P_{\text{gs}}$ histograms to compare with the data in Fig.~\ref{fig:PCorrect}.  The results of these simulations are summarized in those plots.  These simulations did indeed yield the expected general features, specifically the loss of counts from the highest $0.95<P_{\text{gs}}\leq 1$ bin to the broad hump centered about  $P_{\text{gs}}\sim 0.6$ with increasing $T$, and an exponentially suppressed tail at low $P_{\text{gs}}$.  However, the simulations also yielded more small-scale structure in the histograms than was observed in experiment.  

It was hypothesized that slight errors in embedding problem instances in the hardware could have been responsible for the absence of fine structure in the experimental histograms in  Fig.~\ref{fig:PCorrect}.  The PMM elements responsible for nulling small qubit flux offsets are on-chip digital-to-analog flux converters whose least significant bit (LSB) weights were measured to be $\delta\Phi^{\text{LSB}}_q\approx 38\,\mu\Phi_0$.  Therefore, we cannot null qubit degeneracy point offsets $\Phi^0_i$ to better than $\pm\delta\Phi^{\text{LSB}}_q/2$ using PMM.  Furthermore, the PMM elements controlling the coupler and $\iqp$-compensator flux biases, $\Phi^x_{\text{co},ij}$ and $\Phi^x_{I_p,i}$, had LSB weights $\delta\Phi^{\text{LSB}}_{\text{co}}\approx\delta\Phi^{\text{LSB}}_{I_p}\approx 60\,\mu\Phi_0$.  This latter quantity limits the precision to which $K_{ij}\propto M_{ij}$ and $h_i\propto M_i$ can be specified, respectively, thereby distorting problem instances.  We have attempted to simulate the effects of the finite precision of our PMM elements on all 6400 problem instances.  It was observed that such embedding errors only marginally smoothed out the fine structure in the simulated histograms.

A second potential source of error in embedding problem instances onto the hardware was that due to imperfect setting of the CCJJ control biases that are used to null the effects of Josephson junction variations\cite{CCJJ}.  Such imperfections give rise to an apparent flux offset in a qubit that varies monotonically with the annealing control, meaning that $\Phi^0_i\rightarrow \Phi^0_i\left(\Phiccjjx\right)$ in Eq.~(\ref{eqn:hdefn}), as given by Eq.~(B4c) of Ref.~\onlinecite{CCJJ}.  Our single-qubit calibration procedures null $\left|\Phi^0_i\left(\Phi^m_{\text{ccjj}}\right)-\Phi^0_i\left(\Phi^b_{\text{ccjj}}\right)\right|$ to less than $20\,\mu\Phi_0$, where the biases $\Phi^m_{\text{ccjj}}$ and $\Phi^b_{\text{ccjj}}$ straddle the region that is critical for annealing, as depicted in Fig.~\ref{fig:Waveforms}.  Thus, while such residual imperfections will inevitably be present, we nonetheless consider their impact to be subdominant to the errors introduced by the finite precision of the PMM.

Our multiqubit generalized Bloch-Redfield model invokes Ohmic noise spectral densities that extend to high energy, thus providing a mechanism by which each individual qubit can exchange energy with its environment.  Within the context of gate model quantum computation (GMQC), high frequency noise of this type gives rise to the timescale $T_1$ over which a qubit prepared in its excited state would relax towards thermal equilibrium.  In many respects, QA is compromised by the reverse process in which a qubit prepared in its ground state is excited by the environment over a timescale $e^{E_q/k_BT}T_1$, as dictated by the Einstein relation between upward and downward transition rates, where $E_q=\sqrt{\epsilon_q^2+\Delta_q^2}$ is the excitation energy of a qubit for arbitrary bias conditions.  This implies that QA is somewhat more robust against the effects of thermalization than GMQC, although our numerical simulations clearly indicate that the ground state probability $P_{\text{gs}}$ is reduced by these $T_1$-related processes.  Nonetheless, theoretical studies such as Ref.~\onlinecite{BR} have concluded that the computation time need not be severely compromised by the presence of such an environment.

One of the key oversights of our generalized Bloch-Redfield model is the role of non-Ohmic low frequency flux noise, particularly noise spectral densities of the form $S_i(\omega)\propto 1/\omega^{\eta}$ ($\eta>0$) that are commonly referred to as $1/f$ noise.  Within the context of GMQC, low frequency flux noise adiabatically manipulates qubit biases, thus leading to dephasing over a timescale frequently denoted as $T_{\varphi}$.  From the perspective of QA, it is more convenient to envision low frequency flux noise as being responsible for the generation of random flux offsets within each qubit that change with every repetition of an experiment.  These flux offsets will possess Gaussian statistics whose distribution has a width given by the integrated low frequency flux noise $W_{\text{LF}}/2\iqp$ \cite{Martinis}.  According to Eq.~(\ref{eqn:hdefn}), random flux offsets will distort the individual $h_i$ by altering $\Phi^0_i$, thus shifting the intended Hamiltonian off target.   If the random flux offsets are small, one can still expect the hardware to return $\ket{\vec{s}_{\text{gs}}}$ with high $P_{\text{gs}}$, but that probability will differ from predictions made from simulating the system in the absence of low frequency flux noise.  If the random flux offsets are comparable to the applied local flux bias on a given qubit at the critical point during annealing when the population statistics freeze, $W_{\text{LF}}/2\iqp_c\sim h_i M_{\text{AFM}}\iqp_c$, then the hardware could end up `solving' the wrong problem.  Thus, low frequency noise ultimately impacts the {\it precision} to which one can specify a problem instance.  This issue is currently being investigated at a theoretical level.

In terms of a heuristic picture, it appears that QA, when run slowly with respect to the adiabatic limit, progresses as follows: First the system is initialized at an annealing parameter $\Phiccjjx(t=0)=\Phi^m_{\text{ccjj}}$ where the gap between ground and first excited state is much greater than $k_BT$.  The large gap ensures that the ground state is occupied with near certainty.  Next, as $\Phiccjjx$ is slowly lowered the system thermalizes to the environment.  The environment strives to bring the qubit system to a Boltzmann distribution, though the ultimate achievement of that outcome can be hindered by differing transition rates between states.  The system continues to thermalize until the qubit dynamics are significantly impeded by the reduction of $\Delta_q$.  Thereafter, despite the fact that the eigenenergies continue to diverge due to the growth of qubit persistent currents,  the state of the processor will no longer evolve.  Thus the population statistics become frozen at the levels achieved at some higher value of $\Phiccjjx(t)>\Phiccjjx(t_f)=\Phi^b_{\text{ccjj}}$.

To summarize this section of the paper, we have simulated the performance of the coupled eight-qubit system using a generalized Bloch-Redfield model to capture the physics of relaxation processes that couple the individual qubits to an environment in thermal equilibrium.  The amplitude of the white noise spectral density sensed by each qubit was estimated from the widths of macroscopic resonant tunneling peaks.  Comparison of the output of the model to the temperature-dependent probability distributions for three example problems indicated that the model captures the broad features seen in the data.  The success histograms for all 6400 problem instances run on the hardware were roughly reproduced.

\section{Discussion and Open Questions}
\label{sec:discussion}

The experimental results in Sect.~\ref{sec:experiment} and the modeling in Sect.~\ref{sec:simulation} present a compelling physical picture of how the state of our prototype processor evolves when run slowly in comparison to the adiabatic limit.  These results have prompted some interesting conclusions and a series of open questions that we will briefly address below.

Our study has demonstrated that the most probable result from the hardware, when run slowly, is invariably the ground state.  The remaining probability is distributed between low lying states that are roughly within an energy window ${\cal O}(k_BT)$ above the energy of the ground state at the point in the annealing process where the population statistics freeze.  Depending upon the nature of the optimization problem, such low lying states may still constitute acceptable solutions.  This observation suggests two possible modes of running a QA processor in practice:
\begin{itemize}
\item{Run a given optimization problem a statistically large number of times.  Take all $\ket{\vec{s}}$ that are observed with probability greater than the readout resolution threshold and calculate $E(\vec{s})$ using Eq.~(\ref{eqn:QUBO}).  Take whatever $\ket{\vec{s}}$ provides the lowest $E$ as the solution.}
\item{Run a given optimization problem once.  Take the output $\ket{\vec{s}}$ and calculate $E(\vec{s})$ using Eq.~(\ref{eqn:QUBO}).  If $E$ is less than some user-defined threshold, then accept the solution.  If not, then iterate.}
\end{itemize}

\noindent Note that had our processor been operated in the first mode, then it would have returned the correct answer to all 6400 problem instances.

We now turn to the open questions that have been motivated by this work:

{\it What happens when the CCJJ bias ramp time is reduced?}  Given the limited bandwidth of our external bias lines, we were unable to elicit a substantial change in population statistics.  Thus, we were restricted to studying the regime in which the thermalization times were much shorter than the annealing time.  Consequently, we can only claim to have demonstrated QA, not AQO.  Efforts are underway to build a new apparatus that will allow us to probe the regime in which thermalization times exceed the annealing time in an eight-qubit processor.

{\it How does the performance scale with problem size?} The eight-qubit unit cell is too small to be used to address issues concerning scaling.  The intention of this study was to provide a basic demonstration of what we believe to be a complete set of essential ingredients needed for building a scalable QA processor acting in concert.  We will reserve a discussion of the operation of larger portions of a complete 128-qubit chip to a future publication.

{\it How does one implement error correction in QA?}  The work presented in this article is in much the same spirit as that of Ref.~\onlinecite{Yale} - operate a small scale device that `looks' like a basic quantum information processor and run it using the simplest algorithms known.  We have made no attempts beyond statistical sampling to implement any form of error correction in these particular experiments.  This is an active area of research at the moment.

{\it To what precision must the qubit parameters $\iqp$ and $\Delta_q$ be synchronized as a function of $\Phiccjjx(t)$?}  Implicit in many models of AQO is the assumption that all of the qubits are identical throughout the annealing process.  An examination of the data in Fig.~\ref{fig:QubitParameters} reveals that while $\iqp$ is very uniform as a function of annealing parameter $\Phiccjjx$, there are clear discrepancies between values of $\Delta_q$ at the same bias $\Phiccjjx$.  While we have demonstrated that our modest eight-qubit unit cell is capable of solving small-scale ISG problem instances, it is neither clear to what degree the asynchronization of $\Delta_q(t)$ may have impeded its performance nor what the implications of such asynchronization are for large-scale systems.

{\it How do the higher excited states of the} CCJJ rf SQUIDs {\it affect performance?}  To our knowledge, this issue has not been addressed in the literature.  We strongly advise against literal interpretations of many-qubit eigenspectra to energies that exceed that of the second excited state of an rf-SQUID qubit.  This has consequences for studies regarding the ultimate utility of QA. Higher excited states provide fast interwell relaxation mechanisms that could limit failures during a computation.\\
\vspace{-12pt} 

\section{Conclusions}
\label{sec:conclusions}

An eight-qubit unit cell that is part of a superconducting optimization processor has been experimentally investigated.  This processor makes use of several scalable elements, including on-chip programmable flux sources, XY-addressable high fidelity readouts, and the use of a limited number of global analog control lines to provide a variety of time-dependent control signals to multiple devices.  The processor was tested using a large set of randomly generated Ising spin glass problem instances.  The experimental results were shown to be consistent with the predictions of a quantum mechanical theory in which the individual qubits are coupled to a thermal environment.  

We thank J. Hilton, F. Brito, K. Pudenz, S. Han, A. Kleinsasser, and G. Kerber for useful discussions.\\
\vspace{2in}

\pagebreak

\begin{appendix}

\section{Glossary of Abbreviations}
\label{Glossary}

\begin{tabular}{ll}
AFM & antiferromagnetic \\
AQO & adiabatic quantum optimization \\
CCJJ & compound-compound Josephson junction \\
CJJ & compound Josephson junction \\
GMQC & gate model quantum computation \\
ICO & internal coupler \\
IPC & persistent current compensator \\
ISG & Ising spin glass \\
LSB & least significant bit \\
LT & inductance tuner \\
MRT & macroscopic resonant tunneling \\
PMM & programmable magnetic memory \\
QA & quantum annealing \\
QFP & quantum flux parametron \\
QSG & quantum Ising spin glass \\
RO & readout \\
XCO & external coupler\\
1QLZ & single-qubit Landau-Zener\\
2QLZ & two-qubit Landau-Zener\\
\end{tabular}

\section{Example Problem Instance Settings}
\label{InstanceSettings}

\begin{center}
\begin{tabular}{|c|c|c|c|c|} \hline
Parameter & Example A & Example B & Example C \\ \hline\hline
$h_0$ & -1 & -5/7 & -3/7 \\ 
$h_1$ & 4/7 & 6/7 & 2/7 \\ 
$h_2$ & 5/7 & -5/7 & 6/7 \\ 
$h_3$ & 4/7 & -1 & 3/7 \\ 
$h_4$ & -4/7 & 6/7 & -5/7 \\ 
$h_5$ & 1/7 & -3/7 & 2/7 \\ 
$h_6$ & -5/7 & -3/7 & 0 \\ 
$h_7$ & 4/7 & -3/7 & -2/7 \\ 
$K_{04}$ & 2/7 & -6/7 & -5/7 \\ 
$K_{14}$ & 4/7 & -6/7 & 1/7 \\ 
$K_{24}$ & -1/7 & 0 & 0 \\ 
$K_{34}$ & -1/7 & 6/7 & -3/7 \\ 
$K_{05}$ & 4/7 & 6/7 & 5/7 \\ 
$K_{15}$ & -2/7 & 1/7 & 5/7 \\ 
$K_{25}$ & -1/7 & 3/7 & -1/7 \\ 
$K_{35}$ & 6/7 & -1/7 & -4/7 \\ 
$K_{06}$ & 1/7 & 1 & -6/7 \\ 
$K_{16}$ & 1/7 & 3/7 & 5/7 \\ 
$K_{26}$ & 4/7 & -5/7 & -5/7 \\ 
$K_{36}$ & -5/7 & 2/7 & -1/7 \\ 
$K_{07}$ & 6/7 & 1/7 & 1/7 \\ 
$K_{17}$ & 2/7 & 4/7 & -3/7 \\ 
$K_{27}$ & 4/7 & 6/7 & -6/7 \\ 
$K_{37}$ & 3/7 & 0 & 5/7 \\ \hline
\end{tabular}
\end{center}

\end{appendix}


\bibliography{HardwarePerformancePaper}

\begin{thebibliography}{55}
\expandafter\ifx\csname natexlab\endcsname\relax\def\natexlab#1{#1}\fi
\expandafter\ifx\csname bibnamefont\endcsname\relax
  \def\bibnamefont#1{#1}\fi
\expandafter\ifx\csname bibfnamefont\endcsname\relax
  \def\bibfnamefont#1{#1}\fi
\expandafter\ifx\csname citenamefont\endcsname\relax
  \def\citenamefont#1{#1}\fi
\expandafter\ifx\csname url\endcsname\relax
  \def\url#1{\texttt{#1}}\fi
\expandafter\ifx\csname urlprefix\endcsname\relax\def\urlprefix{URL }\fi
\providecommand{\bibinfo}[2]{#2}
\providecommand{\eprint}[2][]{\url{#2}}

\bibitem[{\citenamefont{Kadowaki and Nishimori}(1998)}]{Kadowaki}
\bibinfo{author}{\bibfnamefont{T.}~\bibnamefont{Kadowaki}} \bibnamefont{and}
  \bibinfo{author}{\bibfnamefont{H.}~\bibnamefont{Nishimori}},
  \bibinfo{journal}{Phys. Rev. E} \textbf{\bibinfo{volume}{58}},
  \bibinfo{pages}{5355} (\bibinfo{year}{1998}).

\bibitem[{\citenamefont{Farhi et~al.}(2001)\citenamefont{Farhi, Goldstone,
  Gutmann, Lapan, Lundgren, and Preda}}]{Farhi1}
\bibinfo{author}{\bibfnamefont{E.}~\bibnamefont{Farhi}},
  \bibinfo{author}{\bibfnamefont{J.}~\bibnamefont{Goldstone}},
  \bibinfo{author}{\bibfnamefont{S.}~\bibnamefont{Gutmann}},
  \bibinfo{author}{\bibfnamefont{J.}~\bibnamefont{Lapan}},
  \bibinfo{author}{\bibfnamefont{A.}~\bibnamefont{Lundgren}}, \bibnamefont{and}
  \bibinfo{author}{\bibfnamefont{D.}~\bibnamefont{Preda}},
  \bibinfo{journal}{Science} \textbf{\bibinfo{volume}{292}},
  \bibinfo{pages}{472} (\bibinfo{year}{2001}).

\bibitem[{\citenamefont{Santoro et~al.}(2002)\citenamefont{Santoro,
  Marto\u{n}\={a}k, Tosatti, and Car}}]{Santoro}
\bibinfo{author}{\bibfnamefont{G.~E.} \bibnamefont{Santoro}},
  \bibinfo{author}{\bibfnamefont{R.}~\bibnamefont{Marto\u{n}\={a}k}},
  \bibinfo{author}{\bibfnamefont{E.}~\bibnamefont{Tosatti}}, \bibnamefont{and}
  \bibinfo{author}{\bibfnamefont{R.}~\bibnamefont{Car}},
  \bibinfo{journal}{Science} \textbf{\bibinfo{volume}{295}},
  \bibinfo{pages}{2427} (\bibinfo{year}{2002}).

\bibitem[{\citenamefont{Santoro and Tosatti}(2006)}]{SantoroReview}
\bibinfo{author}{\bibfnamefont{G.~E.} \bibnamefont{Santoro}} \bibnamefont{and}
  \bibinfo{author}{\bibfnamefont{E.}~\bibnamefont{Tosatti}},
  \bibinfo{journal}{J. Phys. A: Math. Gen.} \textbf{\bibinfo{volume}{39}},
  \bibinfo{pages}{R393} (\bibinfo{year}{2006}).

\bibitem[{\citenamefont{Rezakhani et~al.}(2009)\citenamefont{Rezakhani, Kuo,
  Hamma, Lidar, and Zanardi}}]{Lidar1}
\bibinfo{author}{\bibfnamefont{A.~T.} \bibnamefont{Rezakhani}},
  \bibinfo{author}{\bibfnamefont{W.-J.} \bibnamefont{Kuo}},
  \bibinfo{author}{\bibfnamefont{A.}~\bibnamefont{Hamma}},
  \bibinfo{author}{\bibfnamefont{D.~A.} \bibnamefont{Lidar}}, \bibnamefont{and}
  \bibinfo{author}{\bibfnamefont{P.}~\bibnamefont{Zanardi}},
  \bibinfo{journal}{Phys. Rev. Lett.} \textbf{\bibinfo{volume}{103}},
  \bibinfo{pages}{080502} (\bibinfo{year}{2009}).

\bibitem[{\citenamefont{Altshuler
  et~al.}(2009{\natexlab{a}})\citenamefont{Altshuler, Krovi, and
  Roland}}]{Altshuler1}
\bibinfo{author}{\bibfnamefont{B.}~\bibnamefont{Altshuler}},
  \bibinfo{author}{\bibfnamefont{H.}~\bibnamefont{Krovi}}, \bibnamefont{and}
  \bibinfo{author}{\bibfnamefont{J.}~\bibnamefont{Roland}}
  (\bibinfo{year}{2009}{\natexlab{a}}),
  \bibinfo{note}{\texttt{arXiv:0908.2782}}.

\bibitem[{\citenamefont{Farhi et~al.}(2009)\citenamefont{Farhi, Goldstone,
  Gosset, Gutmann, Meyer, and Shor}}]{Farhi2}
\bibinfo{author}{\bibfnamefont{E.}~\bibnamefont{Farhi}},
  \bibinfo{author}{\bibfnamefont{J.}~\bibnamefont{Goldstone}},
  \bibinfo{author}{\bibfnamefont{D.}~\bibnamefont{Gosset}},
  \bibinfo{author}{\bibfnamefont{S.}~\bibnamefont{Gutmann}},
  \bibinfo{author}{\bibfnamefont{H.~B.} \bibnamefont{Meyer}}, \bibnamefont{and}
  \bibinfo{author}{\bibfnamefont{P.}~\bibnamefont{Shor}}
  (\bibinfo{year}{2009}), \bibinfo{note}{\texttt{arXiv:0909.4766}}.

\bibitem[{\citenamefont{Altshuler
  et~al.}(2009{\natexlab{b}})\citenamefont{Altshuler, Krovi, and
  Roland}}]{Altshuler2}
\bibinfo{author}{\bibfnamefont{B.}~\bibnamefont{Altshuler}},
  \bibinfo{author}{\bibfnamefont{H.}~\bibnamefont{Krovi}}, \bibnamefont{and}
  \bibinfo{author}{\bibfnamefont{J.}~\bibnamefont{Roland}}
  (\bibinfo{year}{2009}{\natexlab{b}}),
  \bibinfo{note}{\texttt{arXiv:0912.0746}}.

\bibitem[{\citenamefont{Amin and Choi}(2009)}]{AminAndChoi}
\bibinfo{author}{\bibfnamefont{M.~H.~S.} \bibnamefont{Amin}} \bibnamefont{and}
  \bibinfo{author}{\bibfnamefont{V.}~\bibnamefont{Choi}},
  \bibinfo{journal}{Phys. Rev. A} \textbf{\bibinfo{volume}{80}},
  \bibinfo{pages}{062326} (\bibinfo{year}{2009}).

\bibitem[{\citenamefont{Young et~al.}(2010)\citenamefont{Young, Knysh, and
  Smelyanskiy}}]{Young1}
\bibinfo{author}{\bibfnamefont{A.~P.} \bibnamefont{Young}},
  \bibinfo{author}{\bibfnamefont{S.}~\bibnamefont{Knysh}}, \bibnamefont{and}
  \bibinfo{author}{\bibfnamefont{V.~N.} \bibnamefont{Smelyanskiy}},
  \bibinfo{journal}{Phys. Rev. Lett.} \textbf{\bibinfo{volume}{104}},
  \bibinfo{pages}{020502} (\bibinfo{year}{2010}).

\bibitem[{\citenamefont{Ao and Rammer}(1989)}]{AoRammer}
\bibinfo{author}{\bibfnamefont{P.}~\bibnamefont{Ao}} \bibnamefont{and}
  \bibinfo{author}{\bibfnamefont{J.}~\bibnamefont{Rammer}},
  \bibinfo{journal}{Phys. Rev. Lett.} \textbf{\bibinfo{volume}{62}},
  \bibinfo{pages}{3004} (\bibinfo{year}{1989}).

\bibitem[{\citenamefont{Childs et~al.}(2001)\citenamefont{Childs, Farhi, and
  Preskill}}]{Childs}
\bibinfo{author}{\bibfnamefont{A.~M.} \bibnamefont{Childs}},
  \bibinfo{author}{\bibfnamefont{E.}~\bibnamefont{Farhi}}, \bibnamefont{and}
  \bibinfo{author}{\bibfnamefont{J.}~\bibnamefont{Preskill}},
  \bibinfo{journal}{Phys. Rev. A} \textbf{\bibinfo{volume}{65}},
  \bibinfo{pages}{012322} (\bibinfo{year}{2001}).

\bibitem[{\citenamefont{Sarandy and Lidar}(2005)}]{Sarandy}
\bibinfo{author}{\bibfnamefont{M.~S.} \bibnamefont{Sarandy}} \bibnamefont{and}
  \bibinfo{author}{\bibfnamefont{D.~A.} \bibnamefont{Lidar}},
  \bibinfo{journal}{Phys. Rev. Lett.} \textbf{\bibinfo{volume}{95}},
  \bibinfo{pages}{250503} (\bibinfo{year}{2005}).

\bibitem[{\citenamefont{Roland and Cerf}(2005)}]{Roland}
\bibinfo{author}{\bibfnamefont{J.}~\bibnamefont{Roland}} \bibnamefont{and}
  \bibinfo{author}{\bibfnamefont{N.~J.} \bibnamefont{Cerf}},
  \bibinfo{journal}{Phys. Rev. A} \textbf{\bibinfo{volume}{71}},
  \bibinfo{pages}{032330} (\bibinfo{year}{2005}).

\bibitem[{\citenamefont{Ashhab et~al.}(2006)\citenamefont{Ashhab, Johansson,
  and Nori}}]{Ashab}
\bibinfo{author}{\bibfnamefont{S.}~\bibnamefont{Ashhab}},
  \bibinfo{author}{\bibfnamefont{J.~R.} \bibnamefont{Johansson}},
  \bibnamefont{and} \bibinfo{author}{\bibfnamefont{F.}~\bibnamefont{Nori}},
  \bibinfo{journal}{Phys. Rev. A} \textbf{\bibinfo{volume}{74}},
  \bibinfo{pages}{052330} (\bibinfo{year}{2006}).

\bibitem[{\citenamefont{Tiersch and Sch\"utzhold}(2007)}]{Tiersch}
\bibinfo{author}{\bibfnamefont{M.}~\bibnamefont{Tiersch}} \bibnamefont{and}
  \bibinfo{author}{\bibfnamefont{R.}~\bibnamefont{Sch\"utzhold}},
  \bibinfo{journal}{Phys. Rev. A} \textbf{\bibinfo{volume}{75}},
  \bibinfo{pages}{062313} (\bibinfo{year}{2007}).

\bibitem[{\citenamefont{Fubini et~al.}(2007)\citenamefont{Fubini, Falci, and
  Osterloh}}]{Fubini}
\bibinfo{author}{\bibfnamefont{A.}~\bibnamefont{Fubini}},
  \bibinfo{author}{\bibfnamefont{G.}~\bibnamefont{Falci}}, \bibnamefont{and}
  \bibinfo{author}{\bibfnamefont{A.}~\bibnamefont{Osterloh}},
  \bibinfo{journal}{New J. Phys.} \textbf{\bibinfo{volume}{9}},
  \bibinfo{pages}{134} (\bibinfo{year}{2007}).

\bibitem[{\citenamefont{O'Hara and O'Leary}(2008)}]{OHara}
\bibinfo{author}{\bibfnamefont{M.~J.} \bibnamefont{O'Hara}} \bibnamefont{and}
  \bibinfo{author}{\bibfnamefont{D.~P.} \bibnamefont{O'Leary}},
  \bibinfo{journal}{Phys. Rev. A} \textbf{\bibinfo{volume}{77}},
  \bibinfo{pages}{042319} (\bibinfo{year}{2008}).

\bibitem[{\citenamefont{Amin et~al.}(2008)\citenamefont{Amin, Love, and
  Truncik}}]{Crazy}
\bibinfo{author}{\bibfnamefont{M.~H.~S.} \bibnamefont{Amin}},
  \bibinfo{author}{\bibfnamefont{P.~J.} \bibnamefont{Love}}, \bibnamefont{and}
  \bibinfo{author}{\bibfnamefont{C.~J.~S.} \bibnamefont{Truncik}},
  \bibinfo{journal}{Phys. Rev. Lett.} \textbf{\bibinfo{volume}{100}},
  \bibinfo{pages}{060503} (\bibinfo{year}{2008}).

\bibitem[{\citenamefont{Steffen et~al.}(2003)\citenamefont{Steffen, van Dam,
  Hogg, Breyta, and Chuang}}]{Chuang}
\bibinfo{author}{\bibfnamefont{M.}~\bibnamefont{Steffen}},
  \bibinfo{author}{\bibfnamefont{W.}~\bibnamefont{van Dam}},
  \bibinfo{author}{\bibfnamefont{T.}~\bibnamefont{Hogg}},
  \bibinfo{author}{\bibfnamefont{G.}~\bibnamefont{Breyta}}, \bibnamefont{and}
  \bibinfo{author}{\bibfnamefont{I.}~\bibnamefont{Chuang}},
  \bibinfo{journal}{Phys. Rev. Lett.} \textbf{\bibinfo{volume}{90}},
  \bibinfo{pages}{067903} (\bibinfo{year}{2003}).

\bibitem[{\citenamefont{Kaminsky et~al.}(2004)\citenamefont{Kaminsky, Lloyd,
  and Orlando}}]{Kaminsky1}
\bibinfo{author}{\bibfnamefont{W.~M.} \bibnamefont{Kaminsky}},
  \bibinfo{author}{\bibfnamefont{S.}~\bibnamefont{Lloyd}}, \bibnamefont{and}
  \bibinfo{author}{\bibfnamefont{T.~P.} \bibnamefont{Orlando}}
  (\bibinfo{year}{2004}), \bibinfo{note}{\texttt{arXiv:quant-ph/0403090v2}}.

\bibitem[{\citenamefont{Kaminsky and Lloyd}(2004)}]{Kaminsky2}
\bibinfo{author}{\bibfnamefont{W.~M.} \bibnamefont{Kaminsky}} \bibnamefont{and}
  \bibinfo{author}{\bibfnamefont{S.}~\bibnamefont{Lloyd}}, in
  \emph{\bibinfo{booktitle}{Quantum Computing and Quantum Bits in Mesoscopic
  Systems}} (\bibinfo{publisher}{Kluwer Academic, New York USA},
  \bibinfo{year}{2004}).

\bibitem[{\citenamefont{Grajcar et~al.}(2005)\citenamefont{Grajcar, Izmalkov,
  and Il'ichev}}]{IPHT3QProposal}
\bibinfo{author}{\bibfnamefont{M.}~\bibnamefont{Grajcar}},
  \bibinfo{author}{\bibfnamefont{A.}~\bibnamefont{Izmalkov}}, \bibnamefont{and}
  \bibinfo{author}{\bibfnamefont{E.}~\bibnamefont{Il'ichev}},
  \bibinfo{journal}{Phys. Rev. B} \textbf{\bibinfo{volume}{71}},
  \bibinfo{pages}{144501} (\bibinfo{year}{2005}).

\bibitem[{\citenamefont{Grajcar et~al.}(2006)\citenamefont{Grajcar, Izmalkov,
  van~der Ploeg, Linzen, Plecenik, Wagner, H\"{u}bner, Il'ichev, Meyer, Smirnov
  et~al.}}]{IPHT4Q}
\bibinfo{author}{\bibfnamefont{M.}~\bibnamefont{Grajcar}},
  \bibinfo{author}{\bibfnamefont{A.}~\bibnamefont{Izmalkov}},
  \bibinfo{author}{\bibfnamefont{S.~H.~W.} \bibnamefont{van~der Ploeg}},
  \bibinfo{author}{\bibfnamefont{S.}~\bibnamefont{Linzen}},
  \bibinfo{author}{\bibfnamefont{T.}~\bibnamefont{Plecenik}},
  \bibinfo{author}{\bibfnamefont{T.}~\bibnamefont{Wagner}},
  \bibinfo{author}{\bibfnamefont{U.}~\bibnamefont{H\"{u}bner}},
  \bibinfo{author}{\bibfnamefont{E.}~\bibnamefont{Il'ichev}},
  \bibinfo{author}{\bibfnamefont{H.~G.} \bibnamefont{Meyer}},
  \bibinfo{author}{\bibfnamefont{A.~Y.} \bibnamefont{Smirnov}},
  \bibnamefont{et~al.}, \bibinfo{journal}{Phys. Rev. Lett.}
  \textbf{\bibinfo{volume}{96}}, \bibinfo{pages}{047006}
  (\bibinfo{year}{2006}).

\bibitem[{\citenamefont{Zakosarenko et~al.}(2007)\citenamefont{Zakosarenko,
  Bondarenko, van~der Ploeg, Izmalkov, Linzen, Kunert, Grajcar, Il'ichev, and
  Meyer}}]{IPHTClassicalCircuit}
\bibinfo{author}{\bibfnamefont{V.}~\bibnamefont{Zakosarenko}},
  \bibinfo{author}{\bibfnamefont{N.}~\bibnamefont{Bondarenko}},
  \bibinfo{author}{\bibfnamefont{S.~H.~W.} \bibnamefont{van~der Ploeg}},
  \bibinfo{author}{\bibfnamefont{A.}~\bibnamefont{Izmalkov}},
  \bibinfo{author}{\bibfnamefont{S.}~\bibnamefont{Linzen}},
  \bibinfo{author}{\bibfnamefont{J.}~\bibnamefont{Kunert}},
  \bibinfo{author}{\bibfnamefont{M.}~\bibnamefont{Grajcar}},
  \bibinfo{author}{\bibfnamefont{E.}~\bibnamefont{Il'ichev}}, \bibnamefont{and}
  \bibinfo{author}{\bibfnamefont{H.-G.} \bibnamefont{Meyer}},
  \bibinfo{journal}{Appl. Phys. Lett.} \textbf{\bibinfo{volume}{90}},
  \bibinfo{pages}{022501} (\bibinfo{year}{2007}).

\bibitem[{\citenamefont{Brooke et~al.}(1999)\citenamefont{Brooke, Bitko,
  Rosenbaum, and Aeppli}}]{Brooke1}
\bibinfo{author}{\bibfnamefont{J.}~\bibnamefont{Brooke}},
  \bibinfo{author}{\bibfnamefont{D.}~\bibnamefont{Bitko}},
  \bibinfo{author}{\bibfnamefont{T.~F.} \bibnamefont{Rosenbaum}},
  \bibnamefont{and} \bibinfo{author}{\bibfnamefont{G.}~\bibnamefont{Aeppli}},
  \bibinfo{journal}{Science} \textbf{\bibinfo{volume}{284}},
  \bibinfo{pages}{779} (\bibinfo{year}{1999}).

\bibitem[{\citenamefont{Brooke et~al.}(2001)\citenamefont{Brooke, Rosenbaum,
  and Aeppli}}]{Brooke2}
\bibinfo{author}{\bibfnamefont{J.}~\bibnamefont{Brooke}},
  \bibinfo{author}{\bibfnamefont{T.~F.} \bibnamefont{Rosenbaum}},
  \bibnamefont{and} \bibinfo{author}{\bibfnamefont{G.}~\bibnamefont{Aeppli}},
  \bibinfo{journal}{Nature} \textbf{\bibinfo{volume}{413}},
  \bibinfo{pages}{610} (\bibinfo{year}{2001}).

\bibitem[{\citenamefont{Harris et~al.}(2010)\citenamefont{Harris, Johansson,
  Berkley, Johnson, Lanting, Han, Bunyk, Ladizinsky, Oh, Perminov
  et~al.}}]{CCJJ}
\bibinfo{author}{\bibfnamefont{R.}~\bibnamefont{Harris}},
  \bibinfo{author}{\bibfnamefont{J.}~\bibnamefont{Johansson}},
  \bibinfo{author}{\bibfnamefont{A.~J.} \bibnamefont{Berkley}},
  \bibinfo{author}{\bibfnamefont{M.~W.} \bibnamefont{Johnson}},
  \bibinfo{author}{\bibfnamefont{T.}~\bibnamefont{Lanting}},
  \bibinfo{author}{\bibfnamefont{S.}~\bibnamefont{Han}},
  \bibinfo{author}{\bibfnamefont{P.}~\bibnamefont{Bunyk}},
  \bibinfo{author}{\bibfnamefont{E.}~\bibnamefont{Ladizinsky}},
  \bibinfo{author}{\bibfnamefont{T.}~\bibnamefont{Oh}},
  \bibinfo{author}{\bibfnamefont{I.}~\bibnamefont{Perminov}},
  \bibnamefont{et~al.}, \bibinfo{journal}{Phys. Rev. B}
  \textbf{\bibinfo{volume}{81}}, \bibinfo{pages}{134510}
  (\bibinfo{year}{2010}).

\bibitem[{\citenamefont{Harris et~al.}(2009)\citenamefont{Harris, Lanting,
  Berkley, Johansson, Johnson, Bunyk, Ladizinsky, Ladizinsky, Oh,
  et~al.}}]{CJJCoupler}
\bibinfo{author}{\bibfnamefont{R.}~\bibnamefont{Harris}},
  \bibinfo{author}{\bibfnamefont{T.}~\bibnamefont{Lanting}},
  \bibinfo{author}{\bibfnamefont{A.~J.} \bibnamefont{Berkley}},
  \bibinfo{author}{\bibfnamefont{J.}~\bibnamefont{Johansson}},
  \bibinfo{author}{\bibfnamefont{M.~W.} \bibnamefont{Johnson}},
  \bibinfo{author}{\bibfnamefont{P.}~\bibnamefont{Bunyk}},
  \bibinfo{author}{\bibfnamefont{E.}~\bibnamefont{Ladizinsky}},
  \bibinfo{author}{\bibfnamefont{N.}~\bibnamefont{Ladizinsky}},
  \bibinfo{author}{\bibfnamefont{T.}~\bibnamefont{Oh}}, , \bibnamefont{et~al.},
  \bibinfo{journal}{Phys. Rev. B} \textbf{\bibinfo{volume}{80}},
  \bibinfo{pages}{052506} (\bibinfo{year}{2009}).

\bibitem[{\citenamefont{Berkley et~al.}(2009)\citenamefont{Berkley, Johnson,
  Bunyk, Harris, Johansson, Lanting, Ladizinsky, Tolkacheva, Amin, and
  Rose}}]{QFP}
\bibinfo{author}{\bibfnamefont{A.~J.} \bibnamefont{Berkley}},
  \bibinfo{author}{\bibfnamefont{M.~W.} \bibnamefont{Johnson}},
  \bibinfo{author}{\bibfnamefont{P.}~\bibnamefont{Bunyk}},
  \bibinfo{author}{\bibfnamefont{R.}~\bibnamefont{Harris}},
  \bibinfo{author}{\bibfnamefont{J.}~\bibnamefont{Johansson}},
  \bibinfo{author}{\bibfnamefont{T.}~\bibnamefont{Lanting}},
  \bibinfo{author}{\bibfnamefont{E.}~\bibnamefont{Ladizinsky}},
  \bibinfo{author}{\bibfnamefont{E.}~\bibnamefont{Tolkacheva}},
  \bibinfo{author}{\bibfnamefont{M.~H.~S.} \bibnamefont{Amin}},
  \bibnamefont{and} \bibinfo{author}{\bibfnamefont{G.}~\bibnamefont{Rose}}
  (\bibinfo{year}{2009}), \bibinfo{note}{\texttt{arXiv:0905.0891}}.

\bibitem[{\citenamefont{Johnson et~al.}(2010)\citenamefont{Johnson, Bunyk,
  Maibaum, Tolkacheva, Berkley, Chapple, Harris, Johansson, Lanting, Perminov
  et~al.}}]{PMM}
\bibinfo{author}{\bibfnamefont{M.~W.} \bibnamefont{Johnson}},
  \bibinfo{author}{\bibfnamefont{P.}~\bibnamefont{Bunyk}},
  \bibinfo{author}{\bibfnamefont{F.}~\bibnamefont{Maibaum}},
  \bibinfo{author}{\bibfnamefont{E.}~\bibnamefont{Tolkacheva}},
  \bibinfo{author}{\bibfnamefont{A.~J.} \bibnamefont{Berkley}},
  \bibinfo{author}{\bibfnamefont{E.~M.} \bibnamefont{Chapple}},
  \bibinfo{author}{\bibfnamefont{R.}~\bibnamefont{Harris}},
  \bibinfo{author}{\bibfnamefont{J.}~\bibnamefont{Johansson}},
  \bibinfo{author}{\bibfnamefont{T.}~\bibnamefont{Lanting}},
  \bibinfo{author}{\bibfnamefont{I.}~\bibnamefont{Perminov}},
  \bibnamefont{et~al.}, \bibinfo{journal}{Supercond. Sci. Technol.}
  \textbf{\bibinfo{volume}{23}}, \bibinfo{pages}{065004}
  (\bibinfo{year}{2010}).

\bibitem[{\citenamefont{Barahona}(1982)}]{BarahonaNP}
\bibinfo{author}{\bibfnamefont{F.}~\bibnamefont{Barahona}},
  \bibinfo{journal}{J. Phys. A.: Math. Gen.} \textbf{\bibinfo{volume}{15}},
  \bibinfo{pages}{3241} (\bibinfo{year}{1982}).

\bibitem[{\citenamefont{Barahona}(1994)}]{Barahona}
\bibinfo{author}{\bibfnamefont{F.}~\bibnamefont{Barahona}},
  \bibinfo{journal}{Phys. Rev. B} \textbf{\bibinfo{volume}{49}},
  \bibinfo{pages}{12864} (\bibinfo{year}{1994}).

\bibitem[{\citenamefont{Han et~al.}(1992)\citenamefont{Han, Lapointe, and
  Lukens}}]{CJJ}
\bibinfo{author}{\bibfnamefont{S.}~\bibnamefont{Han}},
  \bibinfo{author}{\bibfnamefont{J.}~\bibnamefont{Lapointe}}, \bibnamefont{and}
  \bibinfo{author}{\bibfnamefont{J.~E.} \bibnamefont{Lukens}},
  \bibinfo{journal}{Phys. Rev. B} \textbf{\bibinfo{volume}{46}},
  \bibinfo{pages}{6338} (\bibinfo{year}{1992}).

\bibitem[{\citenamefont{Orlando et~al.}(1999)\citenamefont{Orlando, Mooij,
  Tian, van~der Wal, Levitov, Lloyd, and Mazo}}]{Orlando}
\bibinfo{author}{\bibfnamefont{T.~P.} \bibnamefont{Orlando}},
  \bibinfo{author}{\bibfnamefont{J.~E.} \bibnamefont{Mooij}},
  \bibinfo{author}{\bibfnamefont{L.}~\bibnamefont{Tian}},
  \bibinfo{author}{\bibfnamefont{C.~H.} \bibnamefont{van~der Wal}},
  \bibinfo{author}{\bibfnamefont{L.~S.} \bibnamefont{Levitov}},
  \bibinfo{author}{\bibfnamefont{S.}~\bibnamefont{Lloyd}}, \bibnamefont{and}
  \bibinfo{author}{\bibfnamefont{J.~J.} \bibnamefont{Mazo}},
  \bibinfo{journal}{Phys. Rev. B} \textbf{\bibinfo{volume}{60}},
  \bibinfo{pages}{15398} (\bibinfo{year}{1999}).

\bibitem[{\citenamefont{Koch et~al.}(2006)\citenamefont{Koch, Keefe, Milliken,
  Rozen, Tsuei, Kirtley, and DiVincenzo}}]{IBM}
\bibinfo{author}{\bibfnamefont{R.~H.} \bibnamefont{Koch}},
  \bibinfo{author}{\bibfnamefont{G.~A.} \bibnamefont{Keefe}},
  \bibinfo{author}{\bibfnamefont{F.~P.} \bibnamefont{Milliken}},
  \bibinfo{author}{\bibfnamefont{J.~R.} \bibnamefont{Rozen}},
  \bibinfo{author}{\bibfnamefont{C.~C.} \bibnamefont{Tsuei}},
  \bibinfo{author}{\bibfnamefont{J.~R.} \bibnamefont{Kirtley}},
  \bibnamefont{and} \bibinfo{author}{\bibfnamefont{D.~P.}
  \bibnamefont{DiVincenzo}}, \bibinfo{journal}{Phys. Rev. Lett.}
  \textbf{\bibinfo{volume}{96}}, \bibinfo{pages}{127001}
  (\bibinfo{year}{2006}).

\bibitem[{\citenamefont{Paauw et~al.}(2009)\citenamefont{Paauw, Fedorov,
  Harmans, and Mooij}}]{Delft}
\bibinfo{author}{\bibfnamefont{F.~G.} \bibnamefont{Paauw}},
  \bibinfo{author}{\bibfnamefont{A.}~\bibnamefont{Fedorov}},
  \bibinfo{author}{\bibfnamefont{C.~J. P.~M.} \bibnamefont{Harmans}},
  \bibnamefont{and} \bibinfo{author}{\bibfnamefont{J.~E.} \bibnamefont{Mooij}},
  \bibinfo{journal}{Phys. Rev. Lett.} \textbf{\bibinfo{volume}{102}},
  \bibinfo{pages}{090501} (\bibinfo{year}{2009}).

\bibitem[{\citenamefont{Cormen et~al.}(2001)\citenamefont{Cormen, Leiserson,
  Rivest, and Stein}}]{graph}
\bibinfo{author}{\bibfnamefont{T.~H.} \bibnamefont{Cormen}},
  \bibinfo{author}{\bibfnamefont{C.~E.} \bibnamefont{Leiserson}},
  \bibinfo{author}{\bibfnamefont{R.~L.} \bibnamefont{Rivest}},
  \bibnamefont{and} \bibinfo{author}{\bibfnamefont{C.}~\bibnamefont{Stein}},
  \emph{\bibinfo{title}{Introduction to Algorithms}} (\bibinfo{publisher}{MIT
  Press}, \bibinfo{year}{2001}).

\bibitem[{\citenamefont{Johansson et~al.}(2009)\citenamefont{Johansson, Amin,
  Berkley, Bunyk, Choi, Harris, Johnson, Lanting, Lloyd, and Rose}}]{LZ}
\bibinfo{author}{\bibfnamefont{J.}~\bibnamefont{Johansson}},
  \bibinfo{author}{\bibfnamefont{M.~H.~S.} \bibnamefont{Amin}},
  \bibinfo{author}{\bibfnamefont{A.~J.} \bibnamefont{Berkley}},
  \bibinfo{author}{\bibfnamefont{P.}~\bibnamefont{Bunyk}},
  \bibinfo{author}{\bibfnamefont{V.}~\bibnamefont{Choi}},
  \bibinfo{author}{\bibfnamefont{R.}~\bibnamefont{Harris}},
  \bibinfo{author}{\bibfnamefont{M.~W.} \bibnamefont{Johnson}},
  \bibinfo{author}{\bibfnamefont{T.~M.} \bibnamefont{Lanting}},
  \bibinfo{author}{\bibfnamefont{S.}~\bibnamefont{Lloyd}}, \bibnamefont{and}
  \bibinfo{author}{\bibfnamefont{G.}~\bibnamefont{Rose}},
  \bibinfo{journal}{Phys. Rev. B} \textbf{\bibinfo{volume}{80}},
  \bibinfo{pages}{012507} (\bibinfo{year}{2009}).

\bibitem[{\citenamefont{Hogg}(2003)}]{3-SAT}
\bibinfo{author}{\bibfnamefont{T.}~\bibnamefont{Hogg}}, \bibinfo{journal}{Phys.
  Rev. A} \textbf{\bibinfo{volume}{67}}, \bibinfo{pages}{022314}
  (\bibinfo{year}{2003}).

\bibitem[{\citenamefont{Childs et~al.}(2002)\citenamefont{Childs, Farhi,
  Goldstone, and Gutmann}}]{MAX-CLIQUE}
\bibinfo{author}{\bibfnamefont{A.}~\bibnamefont{Childs}},
  \bibinfo{author}{\bibfnamefont{E.}~\bibnamefont{Farhi}},
  \bibinfo{author}{\bibfnamefont{J.}~\bibnamefont{Goldstone}},
  \bibnamefont{and} \bibinfo{author}{\bibfnamefont{S.}~\bibnamefont{Gutmann}},
  \bibinfo{journal}{Quantum Inf. Comput.} \textbf{\bibinfo{volume}{2}},
  \bibinfo{pages}{181} (\bibinfo{year}{2002}).

\bibitem[{\citenamefont{Young et~al.}(2008)\citenamefont{Young, Knysh, and
  Smelyanskiy}}]{EXACT-COVER2}
\bibinfo{author}{\bibfnamefont{A.~P.} \bibnamefont{Young}},
  \bibinfo{author}{\bibfnamefont{S.}~\bibnamefont{Knysh}}, \bibnamefont{and}
  \bibinfo{author}{\bibfnamefont{V.~N.} \bibnamefont{Smelyanskiy}},
  \bibinfo{journal}{Phys. Rev. Lett.} \textbf{\bibinfo{volume}{101}},
  \bibinfo{pages}{170503} (\bibinfo{year}{2008}).

\bibitem[{\citenamefont{Izmalkov et~al.}(2004)\citenamefont{Izmalkov, Grajcar,
  Il'ichev, Oukhanski, Wagner, Meyer, Krech, Amin, van~den Brink, and
  Zagoskin}}]{IPHTLZ}
\bibinfo{author}{\bibfnamefont{A.}~\bibnamefont{Izmalkov}},
  \bibinfo{author}{\bibfnamefont{M.}~\bibnamefont{Grajcar}},
  \bibinfo{author}{\bibfnamefont{E.}~\bibnamefont{Il'ichev}},
  \bibinfo{author}{\bibfnamefont{N.}~\bibnamefont{Oukhanski}},
  \bibinfo{author}{\bibfnamefont{T.}~\bibnamefont{Wagner}},
  \bibinfo{author}{\bibfnamefont{H.-G.} \bibnamefont{Meyer}},
  \bibinfo{author}{\bibfnamefont{W.}~\bibnamefont{Krech}},
  \bibinfo{author}{\bibfnamefont{M.~H.~S.} \bibnamefont{Amin}},
  \bibinfo{author}{\bibfnamefont{A.~M.} \bibnamefont{van~den Brink}},
  \bibnamefont{and} \bibinfo{author}{\bibfnamefont{A.~M.}
  \bibnamefont{Zagoskin}}, \bibinfo{journal}{Europhys. Lett.}
  \textbf{\bibinfo{volume}{65}}, \bibinfo{pages}{844} (\bibinfo{year}{2004}).

\bibitem[{\citenamefont{Berns et~al.}(2006)\citenamefont{Berns, Oliver,
  Valenzuela, Shytov, Berggren, Levitov, and Orlando}}]{OliverLandauZener}
\bibinfo{author}{\bibfnamefont{D.~M.} \bibnamefont{Berns}},
  \bibinfo{author}{\bibfnamefont{W.~D.} \bibnamefont{Oliver}},
  \bibinfo{author}{\bibfnamefont{S.~O.} \bibnamefont{Valenzuela}},
  \bibinfo{author}{\bibfnamefont{A.~V.} \bibnamefont{Shytov}},
  \bibinfo{author}{\bibfnamefont{K.~K.} \bibnamefont{Berggren}},
  \bibinfo{author}{\bibfnamefont{L.~S.} \bibnamefont{Levitov}},
  \bibnamefont{and} \bibinfo{author}{\bibfnamefont{T.~P.}
  \bibnamefont{Orlando}}, \bibinfo{journal}{Phys. Rev. Lett.}
  \textbf{\bibinfo{volume}{97}}, \bibinfo{pages}{150502}
  (\bibinfo{year}{2006}).

\bibitem[{\citenamefont{Harris et~al.}(2008)\citenamefont{Harris, Johnson, Han,
  Berkley, Johansson, Bunyk, Ladizinsky, Govorkov, Thom, Uchaikin
  et~al.}}]{MRT}
\bibinfo{author}{\bibfnamefont{R.}~\bibnamefont{Harris}},
  \bibinfo{author}{\bibfnamefont{M.~W.} \bibnamefont{Johnson}},
  \bibinfo{author}{\bibfnamefont{S.}~\bibnamefont{Han}},
  \bibinfo{author}{\bibfnamefont{A.~J.} \bibnamefont{Berkley}},
  \bibinfo{author}{\bibfnamefont{J.}~\bibnamefont{Johansson}},
  \bibinfo{author}{\bibfnamefont{P.}~\bibnamefont{Bunyk}},
  \bibinfo{author}{\bibfnamefont{E.}~\bibnamefont{Ladizinsky}},
  \bibinfo{author}{\bibfnamefont{S.}~\bibnamefont{Govorkov}},
  \bibinfo{author}{\bibfnamefont{M.~C.} \bibnamefont{Thom}},
  \bibinfo{author}{\bibfnamefont{S.}~\bibnamefont{Uchaikin}},
  \bibnamefont{et~al.}, \bibinfo{journal}{Phys. Rev. Lett.}
  \textbf{\bibinfo{volume}{101}}, \bibinfo{pages}{117003}
  (\bibinfo{year}{2008}).

\bibitem[{\citenamefont{Weiss}(1993)}]{Weiss}
\bibinfo{author}{\bibfnamefont{U.}~\bibnamefont{Weiss}},
  \emph{\bibinfo{title}{Quantum Dissipative Systems}}
  (\bibinfo{publisher}{World Scientific, Singapore}, \bibinfo{year}{1993}).

\bibitem[{\citenamefont{Schoelkopf et~al.}(2003)\citenamefont{Schoelkopf,
  Clerk, Girvin, Lehnert, and Devoret}}]{Schoelkopf}
\bibinfo{author}{\bibfnamefont{R.~J.} \bibnamefont{Schoelkopf}},
  \bibinfo{author}{\bibfnamefont{A.~A.} \bibnamefont{Clerk}},
  \bibinfo{author}{\bibfnamefont{S.~M.} \bibnamefont{Girvin}},
  \bibinfo{author}{\bibfnamefont{K.~W.} \bibnamefont{Lehnert}},
  \bibnamefont{and} \bibinfo{author}{\bibfnamefont{M.~H.}
  \bibnamefont{Devoret}}, in \emph{\bibinfo{booktitle}{Quantum Noise in
  Mesoscopic Physics}}, edited by \bibinfo{editor}{\bibfnamefont{Y.~V.}
  \bibnamefont{Nazarov}} \bibnamefont{and}
  \bibinfo{editor}{\bibfnamefont{Y.~M.} \bibnamefont{Blanter}}
  (\bibinfo{publisher}{Springer, New York, USA}, \bibinfo{year}{2003}).

\bibitem[{\citenamefont{Wellstood et~al.}(2004)\citenamefont{Wellstood, Urbina,
  and Clarke}}]{1OverF}
\bibinfo{author}{\bibfnamefont{F.}~\bibnamefont{Wellstood}},
  \bibinfo{author}{\bibfnamefont{C.}~\bibnamefont{Urbina}}, \bibnamefont{and}
  \bibinfo{author}{\bibfnamefont{J.}~\bibnamefont{Clarke}},
  \bibinfo{journal}{Appl. Phys. Lett.} \textbf{\bibinfo{volume}{85}},
  \bibinfo{pages}{5296} (\bibinfo{year}{2004}).

\bibitem[{\citenamefont{Blum}(1981)}]{Blum}
\bibinfo{author}{\bibfnamefont{K.}~\bibnamefont{Blum}},
  \emph{\bibinfo{title}{Density Matrix Theory and Applications}}
  (\bibinfo{publisher}{Plenum Pub. Corp., New York}, \bibinfo{year}{1981}).

\bibitem[{\citenamefont{Amin et~al.}(2009)\citenamefont{Amin, Truncik, and
  Averin}}]{BR}
\bibinfo{author}{\bibfnamefont{M.~H.~S.} \bibnamefont{Amin}},
  \bibinfo{author}{\bibfnamefont{C.~J.~S.} \bibnamefont{Truncik}},
  \bibnamefont{and} \bibinfo{author}{\bibfnamefont{D.~V.}
  \bibnamefont{Averin}}, \bibinfo{journal}{Phys. Rev. A}
  \textbf{\bibinfo{volume}{80}}, \bibinfo{pages}{022303}
  (\bibinfo{year}{2009}).

\bibitem[{\citenamefont{Amin and Averin}(2008)}]{AminAverin}
\bibinfo{author}{\bibfnamefont{M.~H.~S.} \bibnamefont{Amin}} \bibnamefont{and}
  \bibinfo{author}{\bibfnamefont{D.~V.} \bibnamefont{Averin}},
  \bibinfo{journal}{Phys. Rev. Lett.} \textbf{\bibinfo{volume}{100}},
  \bibinfo{pages}{197001} (\bibinfo{year}{2008}).

\bibitem[{\citenamefont{Lanting et~al.}(2009)\citenamefont{Lanting, Berkley,
  Bumble, Bunyk, Fung, Johansson, Kaul, Kleinsasser, Ladizinsky, Maibaum
  et~al.}}]{1OverFGeometry}
\bibinfo{author}{\bibfnamefont{T.}~\bibnamefont{Lanting}},
  \bibinfo{author}{\bibfnamefont{A.~J.} \bibnamefont{Berkley}},
  \bibinfo{author}{\bibfnamefont{B.}~\bibnamefont{Bumble}},
  \bibinfo{author}{\bibfnamefont{P.}~\bibnamefont{Bunyk}},
  \bibinfo{author}{\bibfnamefont{A.}~\bibnamefont{Fung}},
  \bibinfo{author}{\bibfnamefont{J.}~\bibnamefont{Johansson}},
  \bibinfo{author}{\bibfnamefont{A.}~\bibnamefont{Kaul}},
  \bibinfo{author}{\bibfnamefont{A.}~\bibnamefont{Kleinsasser}},
  \bibinfo{author}{\bibfnamefont{E.}~\bibnamefont{Ladizinsky}},
  \bibinfo{author}{\bibfnamefont{F.}~\bibnamefont{Maibaum}},
  \bibnamefont{et~al.}, \bibinfo{journal}{Phys. Rev. B}
  \textbf{\bibinfo{volume}{79}}, \bibinfo{pages}{060509(R)}
  (\bibinfo{year}{2009}).

\bibitem[{\citenamefont{Deppe et~al.}(2007)\citenamefont{Deppe, Mariantoni,
  Menzel, Saito, Kakuyanagi, Tanaka, Meno, Semba, Takayanagi, and
  Gross}}]{Deppe}
\bibinfo{author}{\bibfnamefont{F.}~\bibnamefont{Deppe}},
  \bibinfo{author}{\bibfnamefont{M.}~\bibnamefont{Mariantoni}},
  \bibinfo{author}{\bibfnamefont{E.~P.} \bibnamefont{Menzel}},
  \bibinfo{author}{\bibfnamefont{S.}~\bibnamefont{Saito}},
  \bibinfo{author}{\bibfnamefont{K.}~\bibnamefont{Kakuyanagi}},
  \bibinfo{author}{\bibfnamefont{H.}~\bibnamefont{Tanaka}},
  \bibinfo{author}{\bibfnamefont{T.}~\bibnamefont{Meno}},
  \bibinfo{author}{\bibfnamefont{K.}~\bibnamefont{Semba}},
  \bibinfo{author}{\bibfnamefont{H.}~\bibnamefont{Takayanagi}},
  \bibnamefont{and} \bibinfo{author}{\bibfnamefont{R.}~\bibnamefont{Gross}},
  \bibinfo{journal}{Phys. Rev. B} \textbf{\bibinfo{volume}{76}},
  \bibinfo{pages}{214503} (\bibinfo{year}{2007}).

\bibitem[{\citenamefont{Martinis et~al.}(2003)\citenamefont{Martinis, Nam,
  Aumentado, Lang, and Urbina}}]{Martinis}
\bibinfo{author}{\bibfnamefont{J.~M.} \bibnamefont{Martinis}},
  \bibinfo{author}{\bibfnamefont{S.}~\bibnamefont{Nam}},
  \bibinfo{author}{\bibfnamefont{J.}~\bibnamefont{Aumentado}},
  \bibinfo{author}{\bibfnamefont{K.~M.} \bibnamefont{Lang}}, \bibnamefont{and}
  \bibinfo{author}{\bibfnamefont{C.}~\bibnamefont{Urbina}},
  \bibinfo{journal}{Phys. Rev. B} \textbf{\bibinfo{volume}{67}},
  \bibinfo{pages}{094510} (\bibinfo{year}{2003}).

\bibitem[{\citenamefont{DiCarlo et~al.}(2009)\citenamefont{DiCarlo, Chow,
  Gambetta, Bishop, Johnson, Schuster, Majer, Blais, Frunzio, Girvin
  et~al.}}]{Yale}
\bibinfo{author}{\bibfnamefont{L.}~\bibnamefont{DiCarlo}},
  \bibinfo{author}{\bibfnamefont{J.~M.} \bibnamefont{Chow}},
  \bibinfo{author}{\bibfnamefont{J.~M.} \bibnamefont{Gambetta}},
  \bibinfo{author}{\bibfnamefont{L.~S.} \bibnamefont{Bishop}},
  \bibinfo{author}{\bibfnamefont{B.~R.} \bibnamefont{Johnson}},
  \bibinfo{author}{\bibfnamefont{D.~I.} \bibnamefont{Schuster}},
  \bibinfo{author}{\bibfnamefont{J.}~\bibnamefont{Majer}},
  \bibinfo{author}{\bibfnamefont{A.}~\bibnamefont{Blais}},
  \bibinfo{author}{\bibfnamefont{L.}~\bibnamefont{Frunzio}},
  \bibinfo{author}{\bibfnamefont{S.~M.} \bibnamefont{Girvin}},
  \bibnamefont{et~al.}, \bibinfo{journal}{Nature}
  \textbf{\bibinfo{volume}{460}}, \bibinfo{pages}{240} (\bibinfo{year}{2009}).

\end{thebibliography}

\end{document}